\documentclass[aps,jmp,amsmath,amssymb,twocolumn,10pt,superscriptaddress]{revtex4-2}

\usepackage{graphicx}
\graphicspath{{figures/}}

\usepackage{dcolumn}% Align TABLE columns on decimal point
\usepackage{bm}% bold math
\usepackage[utf8x]{inputenc}
\usepackage{lineno}
\usepackage{lipsum}
\usepackage{resizegather}
\usepackage{amssymb}
\usepackage{amsmath}
\usepackage{booktabs}
\usepackage{xcolor}

\begin{document}

\title{Transverse fluctuations control the assembly of semiflexible filaments}

\author{Valerio Sorichetti}
\affiliation{Institute of Science and Technology Austria, 3400 Klosterneuburg, Austria}
\affiliation{Laboratoire de Physique Théorique et Modèles Statistiques (LPTMS), CNRS, Université Paris-Saclay, F-91405 Orsay, France}

\author{Martin Lenz}
\affiliation{Laboratoire de Physique Théorique et Modèles Statistiques (LPTMS), CNRS, Université Paris-Saclay, F-91405 Orsay, France}
\affiliation{PMMH, CNRS, ESPCI Paris, PSL University, Sorbonne Université,
Université de Paris, F-75005, Paris, France}

%\email{valerio.sorichetti@ist.ac.at}

%%%%%%%%%%%%%%%%%%%%%%%%%%%%%%%%%%%%%%%%%%%%%%%%%
\begin{abstract} 

The kinetics of the assembly of semiflexible filaments through end-to-end annealing is key to the structure of the cytoskeleton, but is not understood. We analyze this problem through scaling theory and simulations, and uncover a regime where filaments ends find each other through bending fluctuations without the need for the whole filament to diffuse. This results in a very substantial speed-up of assembly in physiological regimes, and could help understand the dynamics of actin and intermediate filaments in biological processes such as wound healing and cell division.

\end{abstract}

%%%%%%%%%%%%%%%%%%%%%%%%%%%%%%%%%%%%%%%%%%%%%%%%%
\maketitle

%%%%%%%%%%%%%%%%%%%%%%%%%%%%%%%%%%%%%%%%%%%%
%\ttl{Introduction}

The self-assembly of cytoskeletal filaments is crucial for many cellular functions, including wound healing \cite{abreu2012cytoskeleton},  and cell division \cite{vivante2021dynamics}. The growth kinetics of these filaments strongly influences the morphology of the networks they form, from bundled to entangled structures \cite{kayser2012assembly,falzone2013actin,foffano2016dynamics,schepers2021multiscale,schween2022dual}. Unlike the well-understood actin filaments and microtubules \cite{howard2002mechanics}, intermediate filaments of vimentin and keratin crucial for cell shape and mechanical integrity \cite{sanghvi2017intermediate} mainly grow by end-to-end annealing \cite{colakouglu2009intermediate,winheim2011deconstructing,martin2015model,herrmann2016intermediate,lopez2016lateral}. This mechanism is also at work in worm-like micelles \cite{cates1990statics}, DNA \cite{heinen2019programmable}, some synthetic polymers \cite{flory1946fundamental}, and plays a secondary role in the assembly of actin \cite{sept1999annealing,andrianantoandro2001kinetic} and microtubules \cite{rothwell1986end}. As filaments elongate by end-to-end annealing, their diffusion becomes slower due to an increased viscous drag. The time needed to find other reaction partners then increases, giving rise to \textit{diffusion-limited} growth \cite{berg1985diffusion}. Theoretical models have been proposed to describe the dependence of the polymer growth kinetics on physical properties such as length, flexibility and concentration \cite{wilemski1974diffusion1,wilemski1974diffusion2,sunagawa1975theory,doi1975diffusion,degennes1982kinetics1,degennes1982kinetics2,grosberg1982polymeric}. Many have however employed the Gaussian chain model, which provides a poor description of cytoskeletal filaments \cite{howard2002mechanics}. 

\begin{figure}[b]
\centering
\includegraphics[width=0.99\columnwidth]{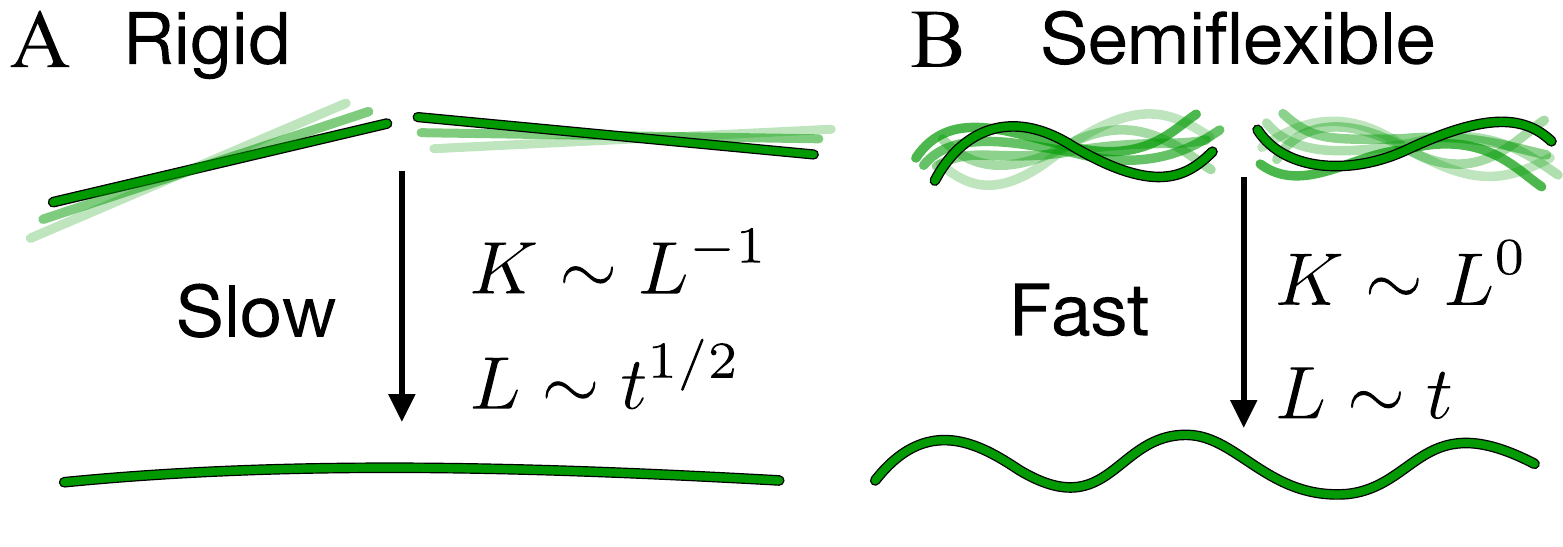}
\caption{Rigid filaments (A) assemble by displacing their center of mass, which results in a reaction rate $K \propto L^{-1}$, with $L$ the mean contour length, and slow growth ($L\propto t^{1/2}$). Here we show that semiflexible filaments (B) take advantage of transverse fluctuations to quickly join their ends, resulting in a constant reaction rate ($K \propto L^0$) and faster growth ($L(t) \propto t$).}
\label{fig:sketch}
\end{figure}

Here, we instead tackle the more general case of semiflexible filaments, and uncover a new assembly regime driven by transverse fluctuations. While rigid rods react slowly due to the need of mobilising the center of mass (Fig.~\ref{fig:sketch}A), these fluctuations speed up the search of bonding partners, leading to faster assembly (Fig.~\ref{fig:sketch}B). We first describe the growth regimes successively encountered by a growing filament, then validate the resulting scaling laws using Brownian dynamics simulations.

%%%%%%%%%%%%%%%%%%%%%%%%%%%%%%%%%%%%%%%%%%%%
%\ttl{Theoretical model}

We model the annealing of semiflexible filaments as an irreversible reaction whereby an $i$-mer and a $j$-mer form an $i+j$-mer (Fig.~\ref{fig:sketch}). The reaction rate constant $K_{i,j}$ generically depends on the lengths of the reacting filaments \cite{berg1985diffusion}. Filaments undergo annealing \textit{via} reactive sites (monomers) located at their ends that bind immediately upon contact, and we assume the system is dilute enough to ignore steric constraints, \textit{e.g.}, entanglements \cite{doi1988theory,lang2018disentangling}. We also neglect hydrodynamic interactions, and describe filament dynamics with the Rouse model \cite{doi1988theory}. Finally, our scaling discussion ignores numerical prefactors as well as length polydispersity, and thus considers a single typical contour length $L$ and reaction rate constant $K(L)$.

The annealing rate of a collection of filaments of length $L$ stems from the dynamics of their reaction sites~\cite{wilemski1974diffusion1,wilemski1974diffusion2,sunagawa1975theory,doi1975diffusion,degennes1982kinetics1,degennes1982kinetics2}. To describe it, we assume a scaling form $x(t) \propto t^\alpha$ for the root-mean squared displacement of one such site. For normal diffusion, $\alpha=1/2$, while $\alpha=1/4$ at short times in a long Gaussian polymer \cite{doi1988theory}. If $\alpha > 1/d$, with $d$ the dimension of space, the monomer explores space in a \textit{non-compact} manner. This means that, if we approximate this exploration as a discrete process in which a site of volume $b^d$ (with $b$ the monomer size) is visited at each step, the number of sites visited at time $t$ is much smaller than $[x(t)/b]^d$. We assume that the reactants are uniformly distributed before the reaction and that the reaction takes place immediately when the reactants come within a distance $\approx b$. Then, in $d=3$ \cite{degennes1982kinetics1}

\begin{equation}
K^{-1} \approx \int_{\tau_b}^{\infty} x^{-3}(t) \ \text{d}t,
\label{eq:k_int}
\end{equation}

\noindent
where $\tau_b$ is the time a monomer takes to move over a distance $b$. In the regimes considered below, this results in $K(L) \propto L^{-\lambda}$, where the exponent $\lambda \geq 0$ depends on the physical process underlying the motion of the reactive sites. The number density of filaments $\nu$ evolves as $\dot \nu = - K(L) \nu^2$. Since $\nu = cb/L$, with $c$ the total monomer density, this implies $L(t) \propto t^{1/(1+\lambda)}$~\cite{vandongen1985dynamic,vandongen1988scaling,meakin1988scaling}.

%%%%%%%%%%%%%%%%%%%%%%%%%%%%%%%%%%%%%%%%%%%%
%\ttl{Theoretical predictions I - rigid rods}

\noindent
Starting from a solution of monomers, filaments are initially much shorter than the persistence length $L_p$ \cite{granek1997semiflexible}, and thus behave as rigid rods ($L_p=\infty$). Their ends undergo diffusive dynamics, \textit{i.e.} $x^2(t) \approx Dt$ where $D$ the center-of-mass diffusion coefficient of the filament. If each monomer is subjected to a viscous friction $\zeta$, we have $D=k_B Tb/ \zeta L$ \cite{doi1988theory}. Equation~\eqref{eq:k_int} with $\tau_b \approx b^2/D$ thus yields $K \approx  b^3 \tau_b^{-1} \approx {k_B Tb^2}/{\zeta L} \approx b^3 \tau^{-1} (b/L)$, where $\tau \approx b^2 \zeta / k_B T$ is the time a free monomer takes to move by $b$. Since $L(t) \propto t^{1/(1+\lambda)}$, the filament length reads

\begin{equation}
 {L(t)} / b \approx \left( {c b^3 t} /{\tau} \right)^{1/2}.
\label{eq:l_regime0}
\end{equation}

\noindent
Thus, both center-of-mass diffusion and filament growth slow down over time.

%%%%%%%%%%%%%%%%%%%%%%%%%%%%%%%%%%%%%%%%%%%%
%\ttl{Theoretical predictions II - fluctuations-driven regime}

As the filaments elongate, bending fluctuations become relevant even as $L \ll L_p$. Indeed, the short-time dynamics of the reactive sites then becomes dominated by bending modes. Their root-mean squared displacement thus grows with time predominantly in the direction perpendicular to the local filament contour \cite{farge1993dynamic,granek1997semiflexible,le2002tracer,huang2014conformations,nikoubashman2016dynamics}. This results in a short-time subdiffusive regime, $x(t) \propto t^{3/8}$. This lasts until the time $\tau_f \approx \tau (L^4/L_p b^3)$ required to relax the longest-wavelength bending mode of the filament. Subsequently, center-of-mass diffusion dominates filament motion. The typical monomer displacement thus reads

\begin{equation}
x(t) \approx
\begin{cases}
\left({b^9}/{L_p}\right)^{1/8} \left(t /{\tau} \right)^{3/8} & \tau \lesssim t \lesssim \tau_f \\
\left({b^3}/L \right)^{1/2}  \left(t/{\tau} \right)^{1/2} & t \gtrsim \tau_f.\\
\end{cases}
\label{eq:x_semiflex}
\end{equation}

\noindent
In the regime considered here, the monomer displacement time $\tau_b$ is computed from the short-time regime of Eq.~\eqref{eq:x_semiflex}, yielding $\tau_b \approx \tau(L_p/b)^{1/3}$. If the total duration $\tau_f$ of the bending-fluctuations-dominated regime is much longer than the monomer displacement time $\tau_b$, this regime dominates the integral of Eq.~\eqref{eq:k_int}, and therefore the reaction rate. We may equivalently require $L\gg L^* \approx b (L_p/b)^{1/3}$. Since $L(t) \propto t^{1/(1+\lambda)}$ , this yields

\begin{equation}
K \approx b^3 \tau_b^{-1} \approx b^3 \tau^{-1} \left({L_p}/b\right)^{-1/3} \qquad (\text{for }L \gg L^*).
\label{eq:k_regime1}
\end{equation}

\noindent
Thus, for filaments longer than $L^*$, the reaction rate is \textit{independent} of $L$, as also found for first-passage problems involving semiflexible filaments \cite{berg1984diffusion,guerin2014cyclization}. A scaling argument leading directly to Eq.~\eqref{eq:k_regime1} is presented in the Supplementary Material. As illustrated in Fig.~\ref{fig:sketch}B, transverse fluctuations then allow the reactive sites to ``find'' each other without center-of-mass motion. As the filaments elongate, their center-of-mass motion slows down, but the short-time dynamics of the reaction sites remains the same. This accounts for the independence of $K$ on $L$ and implies a constant growth speed

\begin{equation}
{L(t)}/b \approx {c b^3 t} /{\tau_b}.
\label{eq:l_regime1}
\end{equation}

\noindent 
Mathematically, this stems from the $\tau_b \lesssim t \lesssim \tau_f$ time domain dominating the integral of Eq.~\eqref{eq:k_int} when $L \gg L^*$.
Equation~\eqref{eq:l_regime1} is valid for $L \gg L^*$, while shorter filaments behave as rigid rods [Eq.~\eqref{eq:l_regime0}]. At the crossover between these two regimes, filaments have a length $L^* \ll L_p$, meaning that bending fluctuations overtake center-of-mass diffusion before the filaments become fully flexible. The crossover time reads $t^*= \tau(c b^3)^{-1} (L_p/b)^{2/3}$.

%%%%%%%%%%%%%%%%%%%%%%%%%%%%%%%%%%%%%%%%%%%%
%\ttl{Theoretical predictions III - Gaussian regime}

As the filaments eventually grow much longer than the persistence length ($L \gg L_p$), the short-time dynamics of the reactive sites is still dominated by the bending modes and independent of $L$ [Eq.~\eqref{eq:x_semiflex}]. At the time $\tilde \tau_f = \tau(L_p/b)^3$, the monomer displacement $x(t)$ becomes of order $L_p$. For later times, the filament behaves as a Gaussian chain \cite{huang2014conformations,nikoubashman2016dynamics} governed by Rouse relaxation modes \cite{doi1988theory}. Segments of the filaments with length $\approx L_p$ then diffuse while elastically coupled with the neighboring segments, leading to a slow, subdiffusive regime $x(t) \propto t^{1/4}$. This  lasts up to the Rouse relaxation time $\tau_R = \tau (L_p L^2/b^3)$. Subsequently, the segments of the chain essentially move together and their displacement is again dominated by center-of-mass diffusion. Combining these three regimes (bending fluctuations, Rouse modes and center-of-mass diffusion), we write for $L \gg L_p$:

\begin{equation}
x(t) \approx
\begin{cases}
 \left({b^9}/{L_p}\right)^{1/8} \left(t/{\tau} \right)^{3/8} & \tau \lesssim t \lesssim \tilde \tau_f \\
L_p \left(t /{\tilde \tau_f}\right)^{1/4} & \tilde \tau_f \lesssim t \lesssim \tau_R\\
(D t)^{1/2} & t\gtrsim \tau_R,
\end{cases}
\label{eq:x_gaussian}
\end{equation}

\noindent
where $D(L)$ is the diffusion constant of the ``rigid rod'' regime. The integral in Eq.~\eqref{eq:k_int} can now be split into three pieces, the last ($t \gtrsim \tau_R$) of which is negligible, yielding

\begin{equation}
K^{-1} \approx  \tau b^{-3}  \left({L_p} /b\right)^{1/3} \left[1 + (3/4) \left( L/{L^{**}}\right)^{1/2} \right],
\end{equation}

\noindent
where $L^{**} = L_p (L_p/b)^{2/3}$ and where each term of the sum stems from one of the remaining pieces of the integral. When $L\gg L^{**}$, the reaction rate thus crosses over from the bending-fluctuations-dominated regime of Eq.~\eqref{eq:k_regime1} to a Gaussian regime with $K \approx b^3 \tau^{-1} (L/L_p)^{-1/2}$. In this regime, the mean contour length increases as

\begin{equation}
{L(t)}/{L^*} \approx \left( {c b^3 t}/{\tau}\right)^{2/3}.
\label{eq:l_regime2}
\end{equation}

\noindent
The crossover time associated with $L^{**}$ is $t^{**}= \tau(c b^3)^{-1} (L_p/b)^2$. This last regime can be understood as follows: After the transverse fluctuations have relaxed ($t>\tilde \tau_f$), the monomers perform a \textit{compact exploration} of space (\textit{i.e.}, densely fill space) and quickly explore the region of size $R \approx L^{1/2}$ occupied by the filaments. The filaments then behave as diffusing reactive spheres with radius $R\propto L^{1/2}$ and diffusion coefficient $D \propto  L^{-1}$. Their reaction rate then obeys the well-known Smoluchowski formula \cite{vandongen1984kinetics}, $K= 4 \pi DR \propto L^{-1/2}$, which results in $L\propto t^{2/3}$\cite{degennes1982kinetics1}. Equation~\eqref{eq:l_regime2} is valid up to $L=L_p^3/b^2$, after which the filament starts to feel its own excluded volume and its dynamics changes \cite{nikoubashman2016dynamics}.

%%%%%%%%%%%%%%%%%%%%%%%%%%%%%%%%%%%%%%%%%%%%
%\ttl{Simulation model}

\begin{figure}[t]
\centering
\includegraphics[width=0.99\columnwidth]{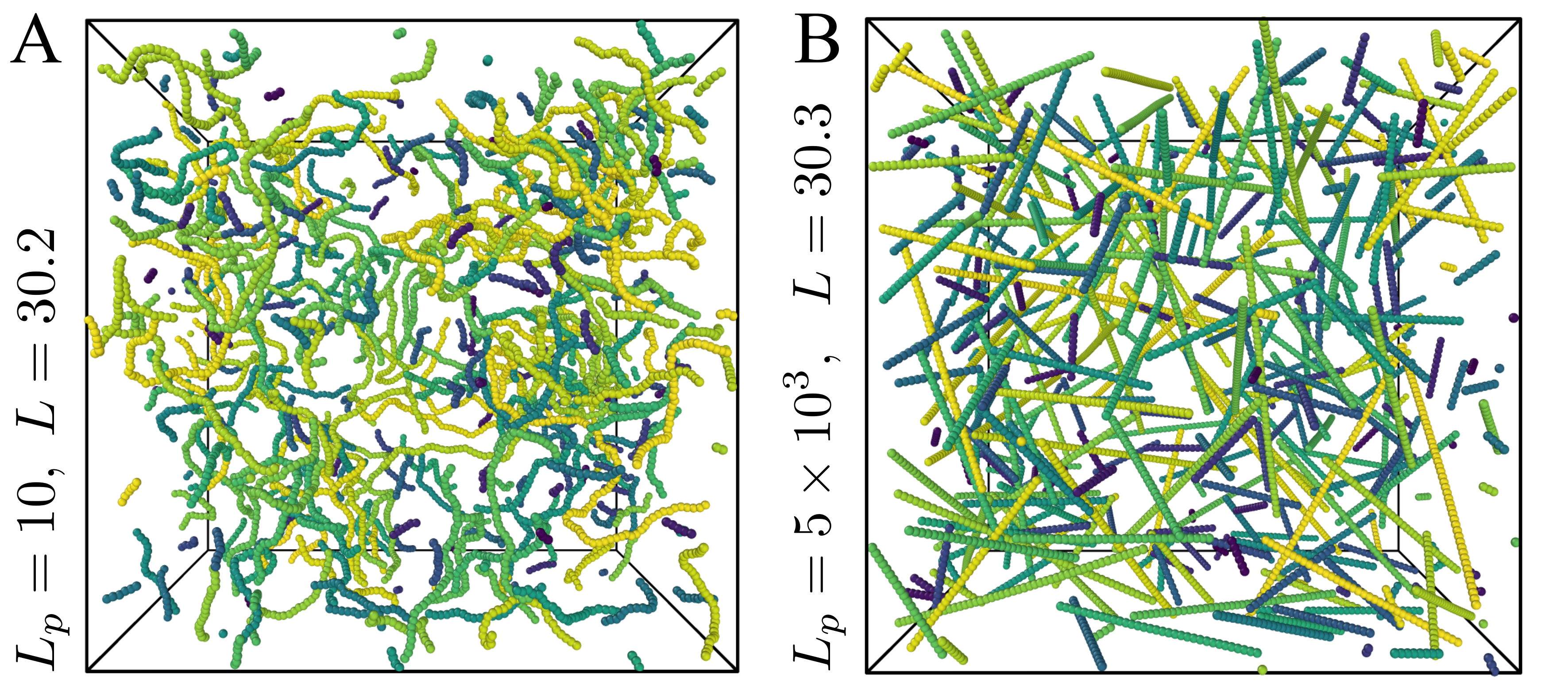}
\caption{Simulation snapshots ($N=8000$ beads) of systems with the same mean contour length $L$ and concentration ($c=10^{-2}$) but with different persistence lengths $L_p$ putting them in the fluctuations-driven (A) and rigid-rod-like (B) regimes. Shorter filaments are colored darker than longer ones.}
\label{fig:snapshots}
\end{figure}

Our scaling results rest on two main assumptions: that the system is characterized by a single typical contour length $L$ and reaction rate constant $K(L)$, and that steric effects can be neglected. To test the robustness of our predictions when these assumptions are relaxed, we run Brownian dynamics simulations of semiflexible polymers undergoing irreversible end-to-end annealing. The polymers are purely repulsive Lennard-Jones beads of diameter $\sigma=1$ connected by finite-extensible nonlinear elastic (FENE) springs~\cite{kremer1990dynamics}. The Lennard-Jones interaction energy is $\epsilon=1$. The system size is $N=8000$ monomers, but we also simulated smaller systems ($N=1000,4096$) to check that there are no significant finite-size effects (Supplementary Material). To simulate semiflexible filaments, we impose an angular potential \cite{svaneborg2020characteristic} $U_\text{ang}(\theta) =  \varepsilon_b [1-\cos(\theta)]$ to bonded triplets, where $\theta$ is the triplet angle and $ \varepsilon_b$ the bending stiffness. For stiff enough filaments $L_p= \varepsilon_b/k_BT$, which we validate by analyzing the bond orientation correlation function (Supplementary Material) and use throughout. We consider $L_p$ values ranging between $10$ and $5 \times 10^3$ (filaments with $L_p<10$ tend to form spurious loops~\cite{panoukidou2022runaway}). To test the validity of our predictions as the concentration $c$ is increased from the dilute to the concentrated regime, we consider $c=10^{-3}, 10^{-2}$ and $10^{-1}$. We note that these values encompass typical ones found for vimentin intermediate filaments in living cells, which are between $0.1$ and a $1$ mg/ml, corresponding to $c$ roughly between $10^{-2}$ and $10^{-1}$~\cite{sivaramakrishnan2008micromechanical,bekker2017optimized}. We carry out the simulations using LAMMPS \cite{plimpton1995fast}, and thermalize the system to an average temperature $k_BT=1.0$ through a Langevin thermostat \cite{schneider1978molecular}. A high monomer friction is imposed in order to simulate Brownian dynamics. To simulate filament annealing, each time two reactive sites come into contact a FENE bond is created between them provided that the angle $\theta$ between prospective bonded triplets is larger than $\theta_\text{min}=160^\circ$ to prevent excessive accumulation of bending energies upon binding. Each monomer can form at most two bonds, so that when polymers are formed, only their ends act as reactive sites. See also the Supplementary Material.

%%%%%%%%%%%%%%%%%%%%%%%%%%%%%%%%%%%%%%%%%%%%
%\ttl{Simulation results}

\begin{figure}[t]
\centering
\includegraphics[width=0.99\columnwidth]{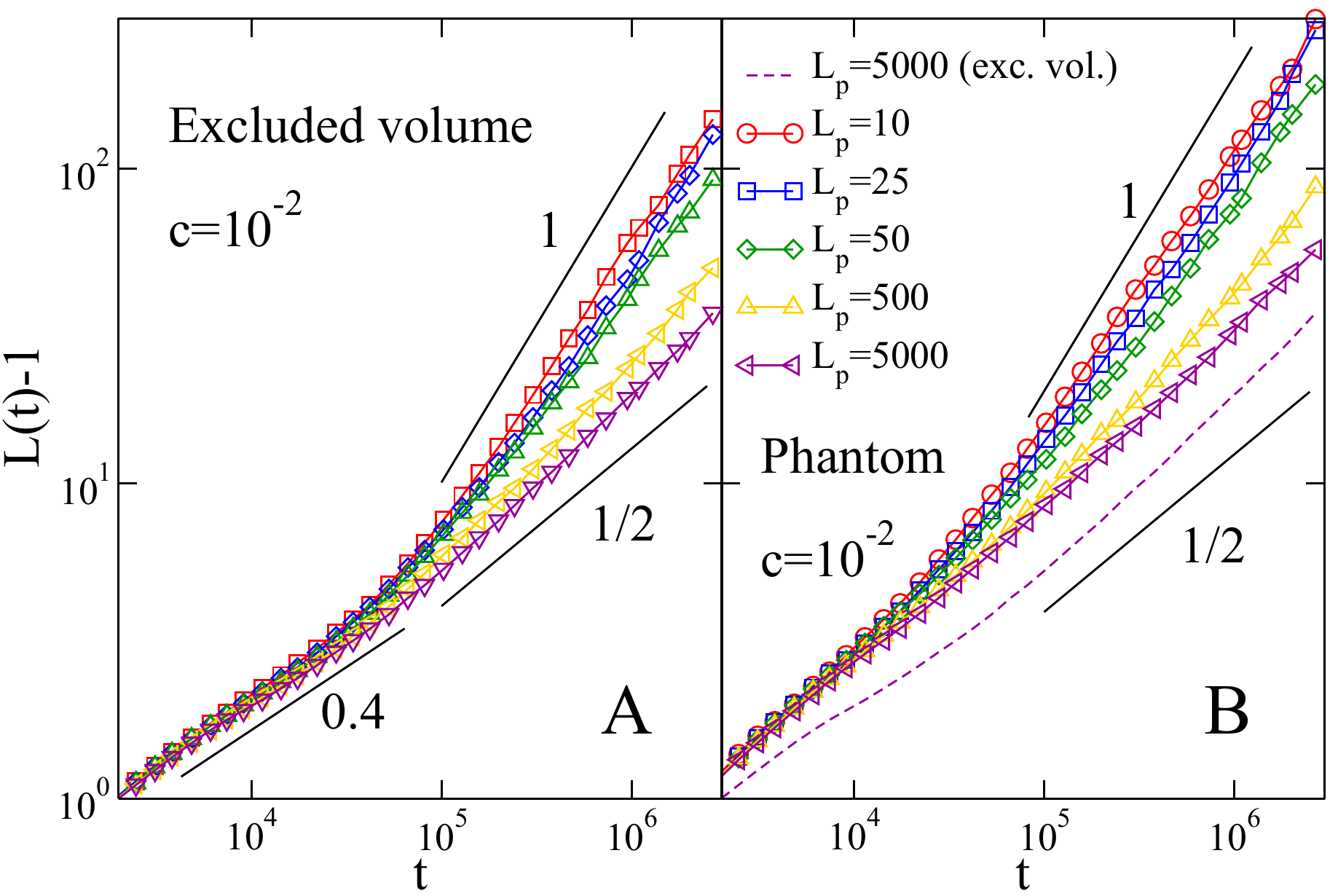}
\caption{Mean contour length as a function of time for excluded volume and phantom filaments with different persistence lengths $L_p$ and concentration $c=10^{-2}$. (A): Excluded volume. (B): Phantom. Dashed line: $L_p=5000$ with excluded volume interactions [same as in panel (A), shown for comparison]. For both systems A and B, the long-time behavior of $L(t)$ follows our predictions for the rigid rods (slope $1/2$) and fluctuations-driven (slope $1$) regimes, depending on $L_p$.}
\label{fig:av_length}
\end{figure}

To assess the validity of our filament annealing dynamics dominated by diffusion and bending fluctuations, we monitor the mean filament contour length $L(t)$ and compare it to our scaling predictions. We start from a monomer solution, implying $L(0)=1$, and thus monitor $L(t)-1$. In Fig.~\ref{fig:av_length}A we show $L(t)$ for systems of polymers with monomer concentration $c=10^{-2}$ and $10 \leq L_p \leq 5 \times 10^3$ (solid lines). At short times, namely for $1 \lesssim L-1\lesssim 3$, we observe a transient regime of sublinear growth $L(t) \propto t^\beta$ with $\beta\simeq 0.4$. We attribute this behavior to slower filament relaxation following binding in the presence of excluded volume interactions (Supplementary Material). After this transient, growth obeys a power law $L(t) \propto t^\beta$ where $\beta$ strongly depends on $L_p$. For large $L_p$, we observe $\beta=1/2$, as predicted for rigid rods.  As $L_p$ is decreased, this exponent increases and approaches $1$ (linear growth) as expected for the fluctuations-dominated regime. 

As filaments elongate, many-body excluded volume interactions become more important and hinder diffusion~\cite{doi1988theory}. This may drastically slow down the motion of the reactive sites, and could conceivably contribute to the observed crossover from sublinear ($\propto t^{1/2}$) to linear growth in Fig.~\ref{fig:av_length}A. To prove that this crossover is instead due to the switching between a rigid rod regime and a fluctuations-dominated one, we simulate a system of ``phantom'' polymers (Fig.~\ref{fig:av_length}B). There, the excluded volume interactions between non-bonded neighbors are removed so that distinct filaments can freely cross each other. The crossover from sublinear to linear growth is preserved in this system, implying that it is not caused by steric effects. There are, however, two differences with Fig.~\ref{fig:av_length}A. First, at very early times $L$ increases approximately as $t^{1/2}$ instead of $t^{0.4}$, suggesting that the transient regime discussed above may be caused by excluded volume effects. Secondly, the phantom polymers display a faster growth ($1.5-2$ times faster for $c=10^{-2}$) both in the sublinear and in the linear regime (see dashed line in Fig.~\ref{fig:av_length}B). To explain this second effect, one could speculate that excluded volume interactions slow down the movement of reactive sites and  thus reduce the prefactor in the $x(t) \propto t^{3/8}$ relation. We however show that this is not the case by directly monitoring the mean-squared displacement of the end monomers of filaments that do not undergo annealing (Supplementary Material). Additionally, we also show that this effect is not due to significant differences in the filament length distribution for phantom and excluded volume filaments (Supplementary Material). This analysis also reveals that filaments that are either much shorter or much longer than $L$ are rare, justifying \textit{a posteriori} our scaling assumption of a single typical length governing the annealing kinetics. We instead attribute the slower assembly in non-phantom systems to the inaccessibility of some potential reaction partners due to steric hindrance~\cite{foffano2016dynamics,tran2022fragmentation}.

\begin{figure}[t]
\centering
\includegraphics[width=0.99\columnwidth]{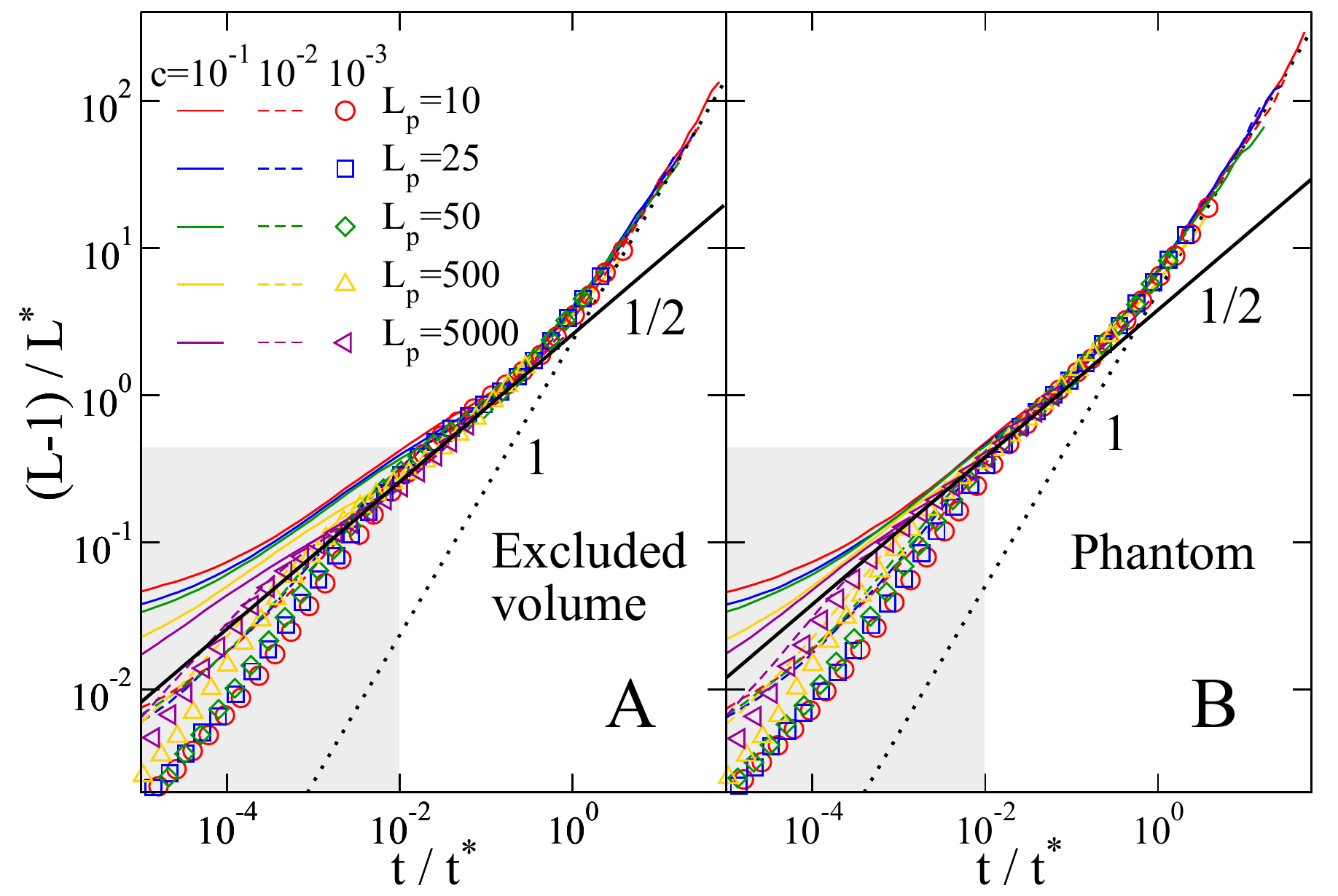}
\caption{Rescaled mean contour length as a function of time for excluded volume and phantom filaments with different persistence lengths $L_p$ and concentrations $c=10^{-3},10^{-2}$ and $10^{-1}$.
Here $L^* = b (L_p/b)^{1/3}$ and $t^* = \tau (cb^3)^{-1} (L_p/b)^{2/3}$. (A): Excluded volume. (B): Phantom. For both systems A and B, the data collapse on a single master curve, in agreement with the theoretical scaling regimes. The density-dependent behavior at small $t$ (shaded area) is due to the rapid formation of a small number of bonds between nearby monomers at the very beginning of the simulation.}
\label{fig:rescaled}
\end{figure}

Finally, to confirm that the main assembly mechanism switches from center-of-mass diffusion to bending fluctuations as filament lengthen, we plot $(L-1)/L^*$ against $t/t^*$, where $L^*$ and $t^*$ are respectively the crossover length and time between the two regimes. Our model predicts that the data should collapse onto the same master curve, with the crossover taking place at $t/t^* \approx 1$, $(L-1)/L^* \approx 1$. We show that this is indeed the case in Fig.~\ref{fig:rescaled}, although the collapse fails for filaments smaller than a dimer ($L \lesssim 2$), where the reaction rate is $K \approx b^3/\tau$ as expected for single monomers \cite{sciortino2008growth}. The collapse there is further distorted by dimerization events occurring within the first simulation time step in the denser regimes ($c \gtrsim 10^{-2}$). Following this initial regime (shaded area in Fig.~\ref{fig:rescaled}), the data collapse on a master curve which displays a crossover between two power-law regimes, confirming our theoretical predictions for both excluded volume and phantom systems. Indeed, we observe a regime with slope $1/2$ (rigid rod regime, dotted line), followed by a rather broad crossover to a linear one (fluctuations-driven regime, solid line). As an additional confirmation of the existence of this crossover, we also measure $K$ directly from the reaction of same-length filaments, finding a good agreement with the scaling prediction (Supplementary Material). While we do not observe the Gaussian regime in our simulations, we recall that our theory predicts its onset only in very long filaments $L \gg L^{**}\propto L_p^{5/3}$. We thus estimate $L^{**}\approx 46$ for our smallest values of $L_p$, which may put this regime out of reach of our current simulations once geometrical prefactors are taken into account.

%%%%%%%%%%%%%%%%%%%%%%%%%%%%%%%%%%%%%%%%%%%%
%\ttl{Discussion and conclusions}

Overall, the fluctuations-driven regime predicted in this work allows for much faster growth of annealing semiflexible filaments compared to rigid rods. This mechanism is likely relevant in the cell cytoskeleton. In vimentin intermediate filaments with $L_p \simeq 1 \ \mu\text{m}$ and $b \simeq 50 \ \text{nm}$ \cite{mucke2004assessing,vicente2022molecular}, we expect our regime to dominate assembly for filament lengths comprised between $L^* \simeq 140 \ \text{nm}$ and $L^{**}\simeq 7.4 \ \mu \text{m}$. This is consistent with the typical lengths between $200  \ \text{nm}$ and $10  \ \mu\text{m}$ observed in cells \cite{robert2016intermediate,terriac2017vimentin}. Estimating the resulting speed-up in assembly as the ratio $K_\text{semiflex}/K_\text{rigid}\approx (L/b)(L_p/b)^{-1/3}$, where $K_\text{rigid}\approx k_BTb^2/\zeta L$ and $K_\text{semiflex}$ is given by Eq.~\eqref{eq:k_regime1} yields a $40$-fold speed-up for a $5\ \mu\text{m}$ vimentin filament.
Actin filaments, which display significant end-to-end annealing under some conditions \cite{sept1999annealing,andrianantoandro2001kinetic}, may be similarly affected. There, $L_p \simeq 18 \ \mu\text{m}$ and $b \simeq 5.5 \ \text{nm}$ \cite{howard2002mechanics} and so $L^* \simeq 82 \ \text{nm}$ and $L^{**} \simeq 4.0 \ \text{mm}$, whereas the \textit{in vivo} filament lengths are comprised between $100 \ \text{nm}$ and a few microns \cite{kasza2010actin}. For a $5\ \mu\text{m}$ actin filament, we estimate a speed-up ratio of $60$. 
Our analysis shows that transverse fluctuations dominate the assembly up to values of $L^{**}$ much longer than the filament persistence length. This implies that the long-length Gaussian regime should very rarely, if ever, be observed. Our findings moreover shed new light on experimental observations of rigid-rod-like assembly kinetics ($K \propto L^{-1}$) in concentrated actin \cite{andrianantoandro2001kinetic} and vimentin \cite{tran2022fragmentation} undergoing annealing \textit{in vitro}. These observations indicate that other phenomena such as lateral interactions (\textit{e.g.} bundling \cite{kayser2012assembly,falzone2013actin,foffano2016dynamics,schepers2021multiscale}), may play a role in these experiments and effectively increase the rigidity of the filaments.

Our numerical simulations reveal that our mechanism does not give rise to widespread filament alignment, and that it is surprisingly robust to molecular crowding and excluded volume interactions. One could indeed naively expect excluded volume effects to significantly slow down network assembly when $L$ becomes comparable with the mesh size $\xi \approx (cb)^{-1/2}$, as would be the case for diffusion in a suspension of rigid rods \cite{doi1988theory}. For a filament volume fraction $c=10^{-2}$ ($c=10^{-1}$), this would lead to significant excluded volume effects for filaments comprising more than $\approx 10$ ($3$) monomers. By contrast, our theory accurately describes the simulated assembly dynamics well beyond these thresholds. This suggests that small-scale end fluctuations remain unhindered by neighboring filaments even in situations where the filament center-of-mass diffusion is largely inhibited, allowing the filaments to keep on annealing. These unhindered fluctuations are evidenced by the preservation of the $x(t)\propto t^{3/8}$ scaling for the filament end displacement even in the presence of excluded volume interactions \cite{lang2018disentangling} (Supplementary Material). This implies that filament assembly continues unabated into the $L>\xi$, ``entangled network'' regime of the semiflexible filament solution, where its short-term elastic modulus and its viscoelastic relaxation time both quickly increase with increasing filament length~\cite{broedersz2014modeling}. In cells, typical values of $\xi$ range roughly between $100$ and $500 \ \text{nm}$ \cite{sivaramakrishnan2008micromechanical,bekker2017optimized}. This corresponds to reduced concentrations $c$ between $10^{-2}$ and $0.25$ for vimentin ($b\simeq 50\,\text{nm}$). This is enough to strongly suppress the filaments' center-of-mass diffusion but not our fluctuations-driven mechanism, implying even larger speed-up ratios than estimated above. The robustness of our assembly mechanism at high concentrations also justifies \textit{a posteriori} neglecting hydrodynamic interactions, as these will be partially screened in concentrated systems~\cite{doi1988theory}. Moreover, even in the dilute regime these interactions only lead to a logarithmic correction to the $x(t)\propto t^{3/8}$ scaling~\cite{granek1997semiflexible,nikoubashman2016dynamics} and we thus do not expect them to significantly alter our predictions. Finally, in our simulations we have considered irreversible bonds and a finite monomer supply. However, knowledge of the annealing rate allows in principle to describe the assembly kinetics also in the presence of severing~\cite{hookway2015microtubule} (if the severing mechanism is known) or equilibrium fragmentation~\cite{tran2022fragmentation}. Moreover, our assembly mechanism is robust with respect to the replenishment of monomers, which can be a relevant process in living cells (Supplementary Material).

Our estimates thus suggest that the mechanism described here may be crucial in allowing the cell to quickly assemble cytoskeletal structures in response to external stimuli. Beyond questions of time scales, these considerations may shift the balance between filament growth and, \emph{e.g.}, bundling or the build-up of entanglements during nonequilibrium cytoskeletal self-assembly. Indeed, It has been shown both in actin~\cite{falzone2012assembly,foffano2016dynamics} and intermediate filaments~\cite{kayser2012assembly,schepers2021multiscale} that differences in filament growth kinetics can lead to networks with markedly different mesh size, bundle density/diameter and mechanical properties. Thus, the mechanism of growth kinetics is likely to have a profound impact on dictating the very structure and mechanics of cytoskeletal networks.

%%%%%%%%%%%%%%%%%%%%%%%%%%%%%%%%%%%%%%%%%%%%%%%%%%%%%%%%%
\acknowledgments
%%%%%%%%%%%%%%%%%%%%%%%%%%%%%%%%%%%%%%%%%%%%%%%%%%%%%%%%%

The authors thank Cécile Leduc and Duc-Quang Tran for invaluable help with understanding the experimental behavior of intermediate filaments, and Raphael Voituriez, Nicolas Levernier and Alexander Grosberg for fruitful discussion on the theoretical model. V.~S. also thanks Davide Michieletto, Maria Panoukidou and Lorenzo Rovigatti for very helpful suggestions on the simulation model.  M.~L. 	was supported by Marie Curie Integration Grant PCIG12-GA-2012-334053, “Investissements d’Avenir” LabEx PALM (ANR-10-LABX- 0039-PALM), ANR grants ANR-15-CE13-0004-03, 	ANR-21-CE11-0004-02 and ANR-22-CE30-0024, as well as ERC Starting 	Grant 677532. M.~L.’s group belongs to the CNRS consortium AQV. Part of this work was performed using HPC resources from GENCI–IDRIS (Grants 2020-A0090712066 and 2021-A0110712066).

%%%%%%%%%%%%%%%%%%%%%%%%%%%%%%%%%%%%%%%%%%%%%%%%%
\bibliography{bibliography.bib}

\end{document}

%\title{Transverse fluctuations control the assembly of semiflexible filaments \protect\\ Supplementary Information}

\title{Transverse fluctuations control the assembly of semiflexible filaments -- Supplementary Information}

\author{Valerio Sorichetti}
\affiliation{Institute of Science and Technology Austria, 3400 Klosterneuburg, Austria}
\affiliation{Laboratoire de Physique Théorique et Modèles Statistiques (LPTMS), CNRS, Université Paris-Saclay, F-91405 Orsay, France}

\author{Martin Lenz}
\affiliation{Laboratoire de Physique Théorique et Modèles Statistiques (LPTMS), CNRS, Université Paris-Saclay, F-91405 Orsay, France}
\affiliation{PMMH, CNRS, ESPCI Paris, PSL University, Sorbonne Université,
Université de Paris, F-75005, Paris, France}

\maketitle

\setcounter{equation}{0}
\setcounter{figure}{0}
\setcounter{table}{0}
\setcounter{section}{0}

\renewcommand{\theequation}{S\arabic{equation}}
\renewcommand{\thefigure}{S\arabic{figure}}
\renewcommand{\thetable}{S\arabic{table}}
\renewcommand{\thesection}{S\Roman{section}}

%%%%%%%%%%%%%%%%%%%%%%%%%%%%%%%%%%%%%%%%%%%%%%%%%%%%%%%%%
\section{Scaling argument for the fluctuation-driven growth regime} \label{sec:si-scaling}
%%%%%%%%%%%%%%%%%%%%%%%%%%%%%%%%%%%%%%%%%%%%%%%%%%%%%%%%%

In this section, we present a simple scaling argument to derive Eq.~(4). Let us consider two semiflexible filaments of length $L$ with diffusion coefficient $D$, and let us consider a spherical region whose center is the center of mass of the filaments and whose radius is $L/2$. When the two filaments meet by diffusing, the respective spherical regions, of volume $V_c \approx L^3$ (neglecting numerical prefactors), overlap for a time $\tau_c \approx L^2/D$. During this time, the reactive ends of the two filaments perform a non-compact exploration of space, with their root-mean squared displacement given by $x(t) \approx \left({b^9}/{L_p}\right)^{1/8} \left(t/{\tau} \right)^{3/8}$, with $\tau = b^3/LD$. Thus, the time needed to explore a volume $b^3$ is $\tau_b = (b^3/LD)(L_p/b)^{1/3}$. Accordingly, the time required to explore the whole volume $V_c$ is $\tau_e \approx \tau_b (V_c/b^3) \approx (L^2/D)(L_p/b)^{1/3}$. Thus, over the time $\tau_c$, only a fraction $\tau_c/\tau_e \approx (L^2/D) (L_p/b)^{-1/3}$ of the volume $V_c$ will be explored. This makes use of the fact that the reactive ends perform a non-compact exploration of space. If a similar argument was repeated for Gaussian polymers, for which the ends perform a \textit{compact} exploration of space, one would find that the whole volume $V_c$ is explored during the time $\tau_c$. Finally, the reaction rate can be obtained as $K \approx V_c/\tau_e \approx LD (L_p/b)^{-1/3}$, which is equivalent to Eq.~(4).

%%%%%%%%%%%%%%%%%%%%%%%%%%%%%%%%%%%%%%%%%%%%%%%%%%%%%%%%%
\section{Simulation model}
\label{sec:model}
%%%%%%%%%%%%%%%%%%%%%%%%%%%%%%%%%%%%%%%%%%%%%%%%%%%%%%%%%

We run $NVT$ Brownian dynamics simulations of a system of $N=8000$ particles (monomers) in a cubic box with periodic boundary conditions. The number density of the monomers is $c=N/V$, where $V$ is the system's volume. We consider here $c=10^{-3},10^{-2}$ and $10^{-1}$. The monomers interact thorough the purely repulsive WCA potential \cite{weeks1971role}, a version of the Lennard-Jones potential which is cut and shifted at its minimum to model excluded volume interactions:

\begin{equation}
U_\text{\footnotesize{WCA}} (r) = 
\begin{cases}
4 \epsilon \left[ \left(\frac \sigma r\right)^{12} -\left(\frac \sigma r\right)^6+ \frac 1 4 \right] & r \leq 2^{1/6} \sigma\\
0 & \text{otherwise}.\\
\end{cases}
\label{eq:wca}
\end{equation}

\noindent
Bonded monomers additionally interact through a finite-extensible-nonlinear-elastic (FENE) potential,

\begin{equation}
U_{\text{FENE}}(r)= -\frac  {K r_0^2} 2 \ln \left[1-(r/r_0)^2\right],
\label{eq:fene}
\end{equation}

\noindent
with $K=30$ and $r_0=1.5$ (Kremer-Grest model \cite{kremer1990dynamics}). These values are chosen in such a way to prevent chain crossing. Since non-bonded interactions are purely
repulsive, this model mimics the behavior of polymers in an athermal
solvent \cite{doi1988theory}. Here we use reduced units, so that $\sigma=1$, $\epsilon=1$, $k_B=1$ (Boltzmann's constant), and the unit mass $m$ is the monomer's mass. The units of temperature, number density and
time are respectively $[T]=\epsilon/k_B$ $[c]=
\sigma^{-3}$ and $[t]=\sqrt{m \sigma^2/\epsilon}$. 

\begin{figure}[t]
\centering
\includegraphics[width=0.45\textwidth]{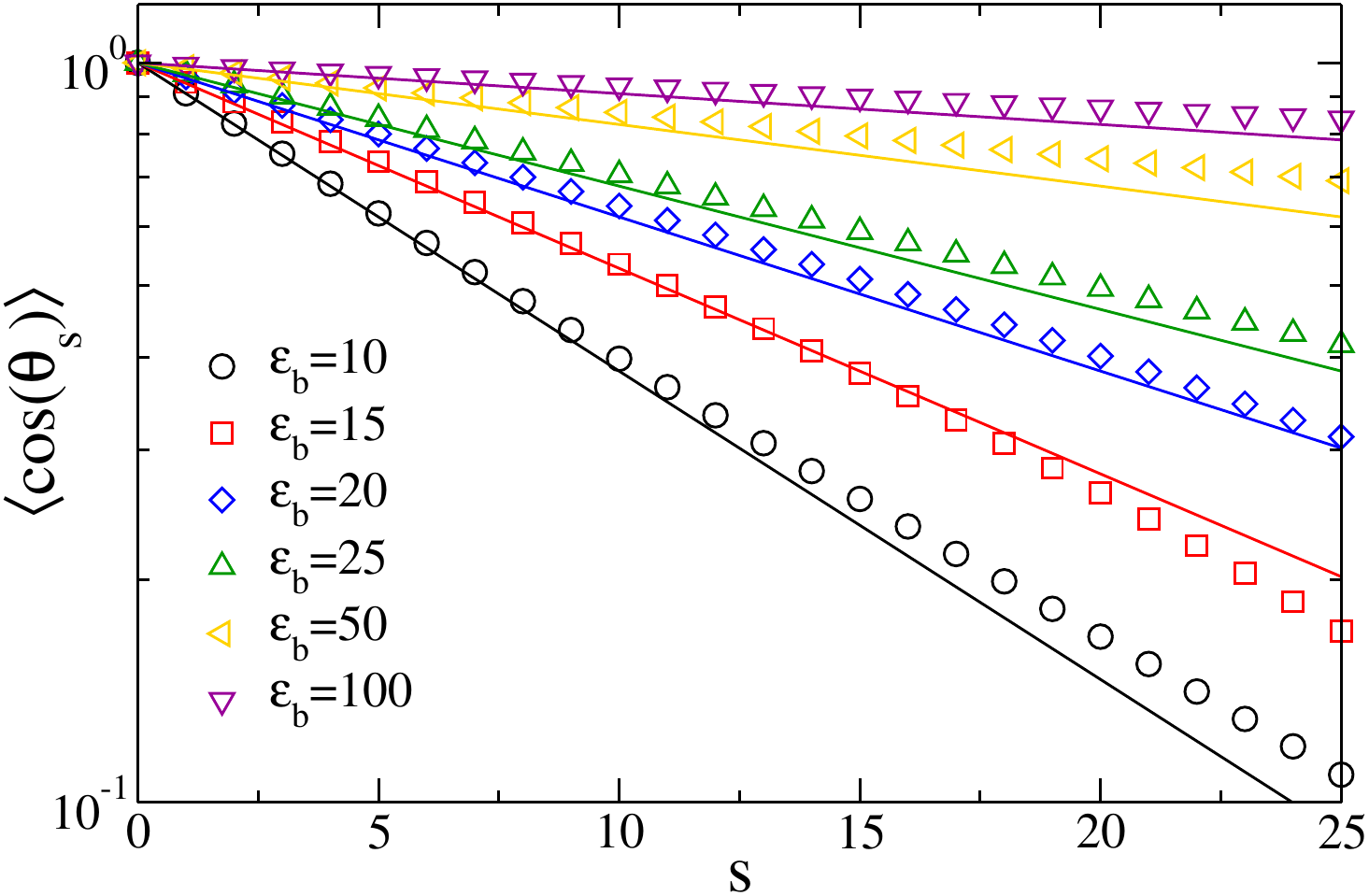}
\caption{Bond orientation correlation function, Eq.~\eqref{eq:bond_corr_def}, obtained from simulations of polymers with different bending energies $\varepsilon_b$, compared to the theoretically predicted exponential decay of Eq.~\eqref{eq:bond_corr_exp} (solid lines).}
\label{fig:bond_corr}
\end{figure}

In addition to the WCA and FENE potentials, bonded triplets interact through a bending potential that allows us to tune chain stiffness \cite{svaneborg2020characteristic}, 

\begin{equation}
U_\text{ang}(\theta) =  \varepsilon_b [1-\cos(\theta)],
\end{equation}

\noindent where $\theta$ is the triplet angle and $ \varepsilon_b$ is the bending stiffness. For stiff enough polymers, $L_p= \varepsilon_b/T$, and thus we define here $L_p$ using this relation. The validity of this relation was also confirmed by analyzing the bond orientation correlation function $\langle \cos(\theta_s)\rangle$, defined as \cite{wittmer2004long}

\begin{equation}
\langle \cos(\theta_s) \rangle \equiv  \left\langle \frac{\mathbf b_k \cdot \mathbf b_{k+s}}{|\mathbf b_k| \ |\mathbf b_{k+s}|} \right\rangle,
\label{eq:bond_corr_def}
\end{equation}

\noindent
where $\mathbf b_k \equiv\mathbf r_{k+1} - \mathbf r_k$ is the $k$-th bond vector, $\langle \rangle$ denote ensemble averages taken over all bond vectors separated by a chemical distance $s$. The persistence length $L_p$ of the polymers can be estimated by the exponential decay of this correlation function \cite{hsu2016static}:

\begin{equation}
\langle \cos(\theta_s)\rangle \propto e^{-sb/L_p} = e^{-sbT/\varepsilon_b},
\label{eq:bond_corr_exp}
\end{equation}

\noindent where $b\simeq 0.96$ is the bond length. As shown in Fig.~\ref{fig:bond_corr}, by comparing $\langle \cos(\theta_s)\rangle$ obtained from simulations to the prediction of Eq.~\eqref{eq:bond_corr_def}, we have verified that the relation $L_p= \varepsilon_b/T$ is satisfied for all $\varepsilon_b\geq 10$, which is the smallest value considered here. We also note that for $L_p < 10$ looping of the polymers (cyclization) is not negligible \cite{panoukidou2022runaway} and leads to a different assembly kinetics, as chains that have formed loops are not reactive.

The solvent is simulated implicitly using a Langevin thermostat \cite{schneider1978molecular}, so that the dynamics of each particle is governed by the following equation:

\begin{equation}
m \ddot{\bm{r}} = -\zeta \dot{\bm{r}} - \bm{\nabla} U + \bm{\eta} (t),
\end{equation}

\noindent
where $\bm{r}$ is the position of a particle, $\zeta$ the viscous friction it experiences, $m$ its mass, and $U$ its potential energy. The term $\boldsymbol{\eta}$ is a stochastic force which represents the collisions with solvent molecules, and satisfies $\langle \boldsymbol{\eta}(t) \rangle = 0$ and $\langle \eta_\alpha (t) \eta_\beta (t') \rangle =  {2m \zeta k_B T \delta_{\alpha,\beta} \delta(t-t')}$, with $\eta_{\alpha}$ its spatial components. To simulate Brownian dynamics, we choose a high friction coefficient, $\zeta = 200$, so that the diffusion time of a free monomer is $\tau=\sigma^2\zeta / k_B T=200$. The Langevin thermostat keeps the average temperature of the system constant at $T=1$. The equation of motions are integrated using the velocity Verlet algorithm with a time step $\delta t = 10^{-3}$. The initial state of the system is a monodisperse solution of monomers, so that $L(0)=1$. For each simulations, we perform two independent realizations with different initial conditions. The data for the mean filament length $L(t)$ reported in the main text are averaged over these two independent realizations.

When the distance $r$ between two unbonded monomers satisfies $r<r_\text{bond} = 1.122 \simeq 2^{1/6}$ (minimum of the WCA potential), provided that each of the two monomers have fewer than two bonded neighbors, a FENE bond is created between them. We note that, since the reaction happens instantaneously as soon as $r<r_\text{bond}$, a small number of reactions take place during the first time step, \textit{i.e.}, $L(\delta t) > 1$. We impose the additional condition on bonding that the angle $\theta$ between prospective bonded triplets must be larger than $\theta_\text{min}=160^\circ$ to prevent excessive accumulation of bending energy as a result of the bond formation. This choice is also in agreement with recent experimental results, which suggest that intermediate polymer annealing can only take place if there is a high degree of local alignment between the reacting filaments \cite{tran2022fragmentation}. We have also tested smaller values of $\theta_\text{min}$, down to $\theta_\text{min}=140^\circ$ verifying that there is no qualitative difference in the observables studied in this work (not shown). 

We note that, despite the precautions taken to avoid the formation of loops, these are occasionally (although very rarely) formed. Sometimes, this is due to a  filament binding to itself across the periodic boundary conditions. This rare occurrence is more common for the highest density here considered ($c=10^{-1}$) and for phantom filaments, whereas the formation of loops that do not cross the periodic boundaries is more common at the lowest density ($c=10^{-3}$). Since the fraction of monomers which are part of loops is always $<1.5\%$ for $c=10^{-2}$, and $<6\%$ for $c=10^{-3}$ and $c=10^{-1}$, we do not expect their presence to significantly alter our results.

%%%%%%%%%%%%%%%%%%%%%%%%%%%%%%%%%%%%%%%%%%%%%%%%%%%%%%%%%
\section{Finite-size effects analysis}
%%%%%%%%%%%%%%%%%%%%%%%%%%%%%%%%%%%%%%%%%%%%%%%%%%%%%%%%%

\begin{figure}[h]
\centering
\includegraphics[width=0.45\textwidth]{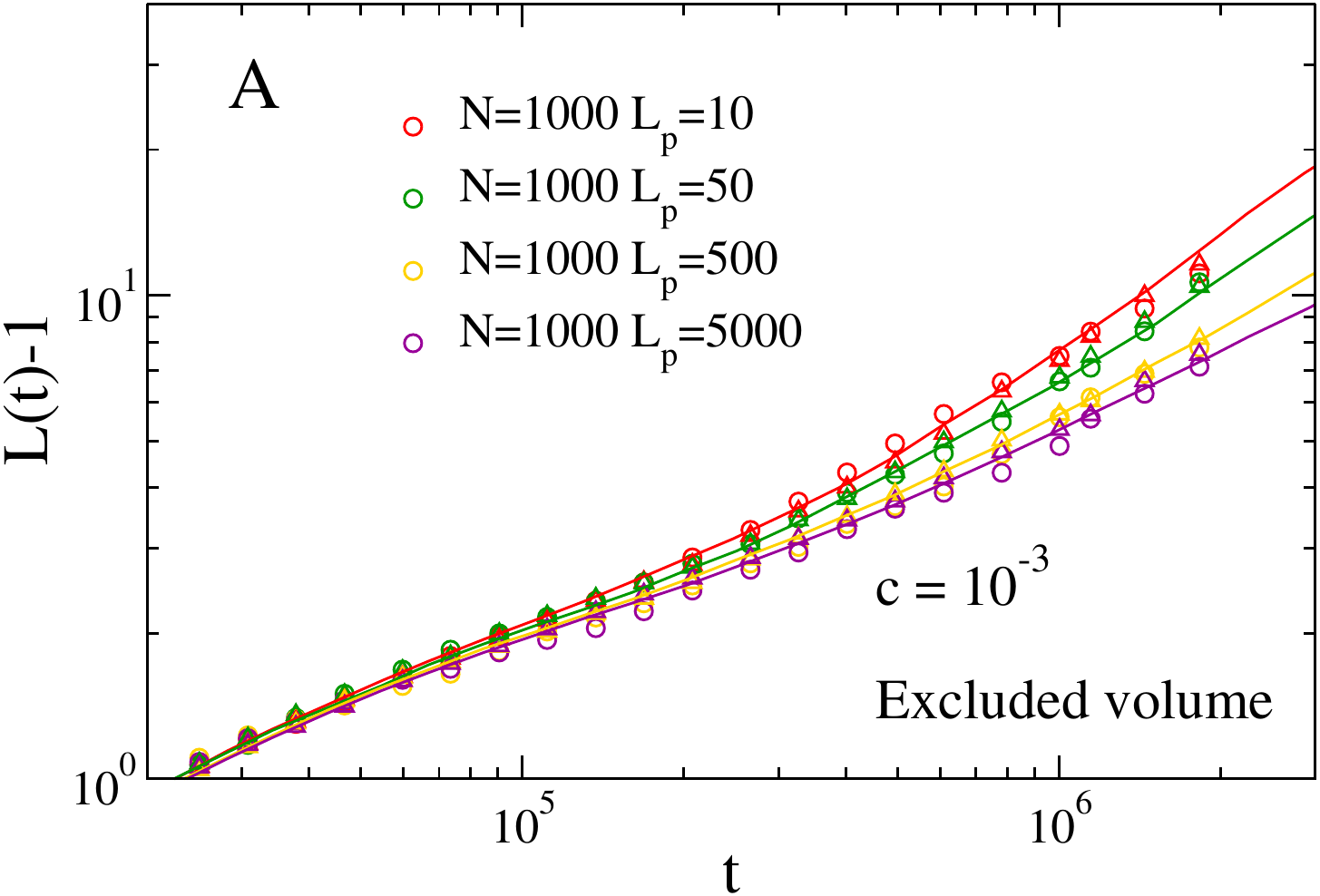}
\includegraphics[width=0.45\textwidth]{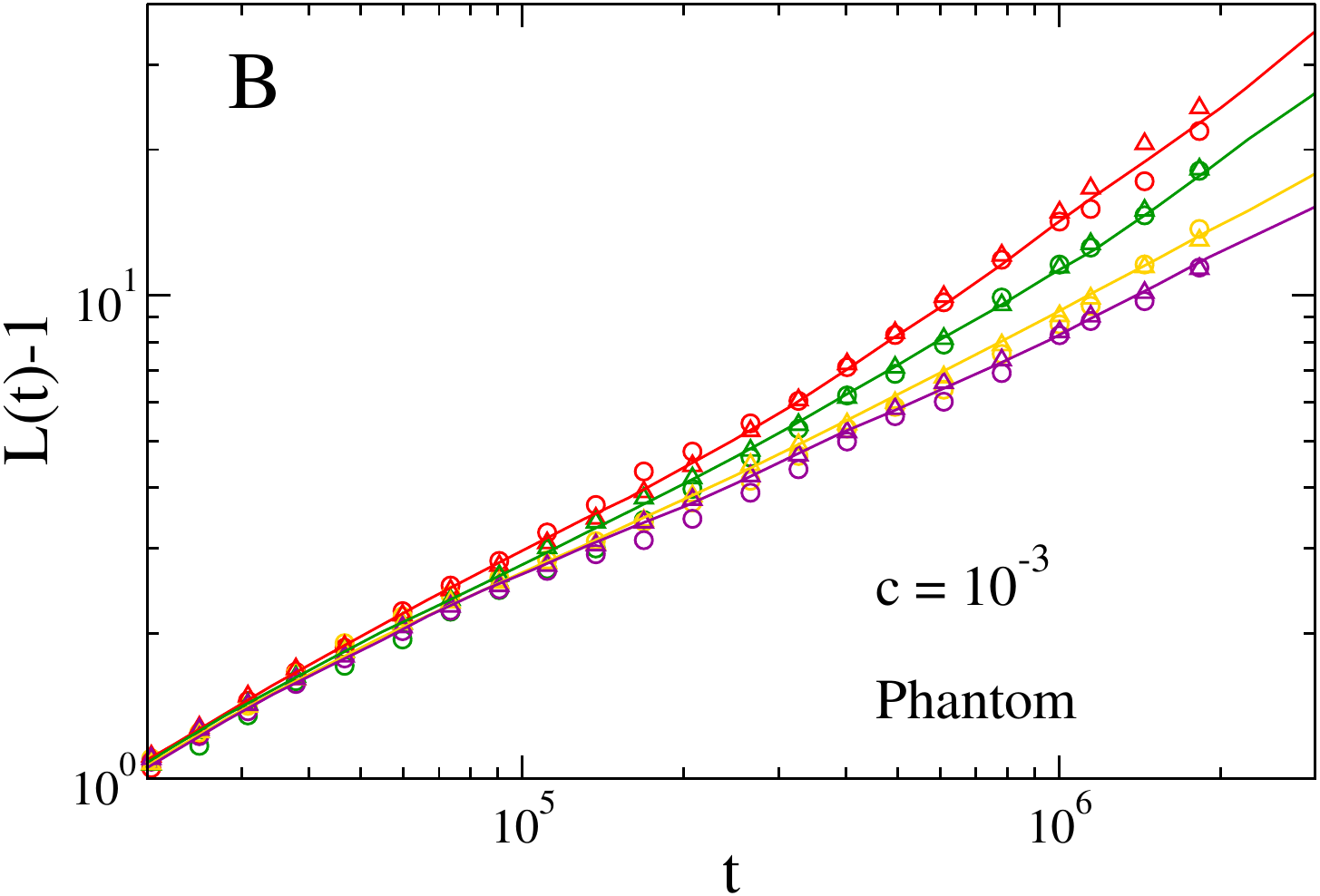}
\includegraphics[width=0.45\textwidth]{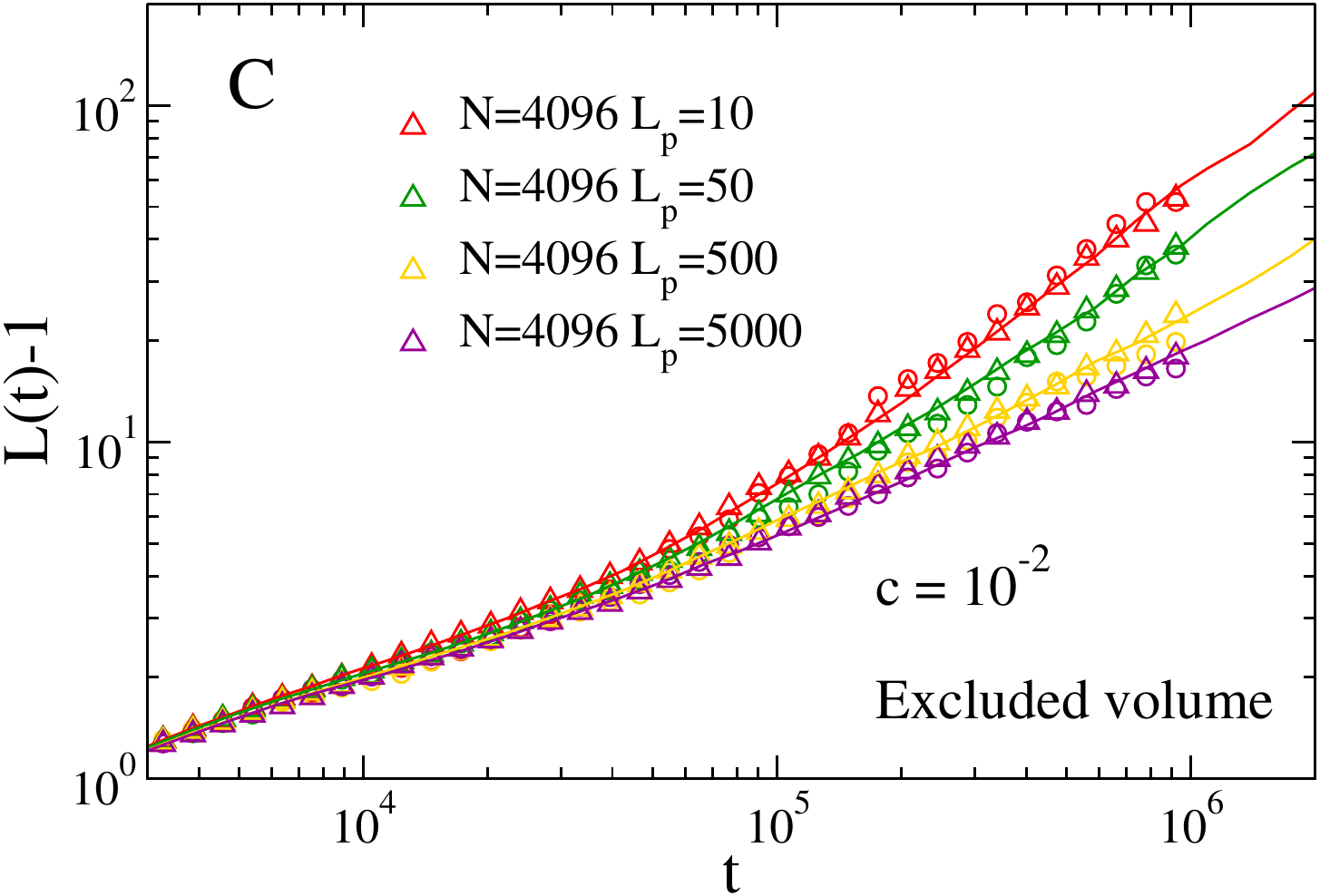}
\includegraphics[width=0.45\textwidth]{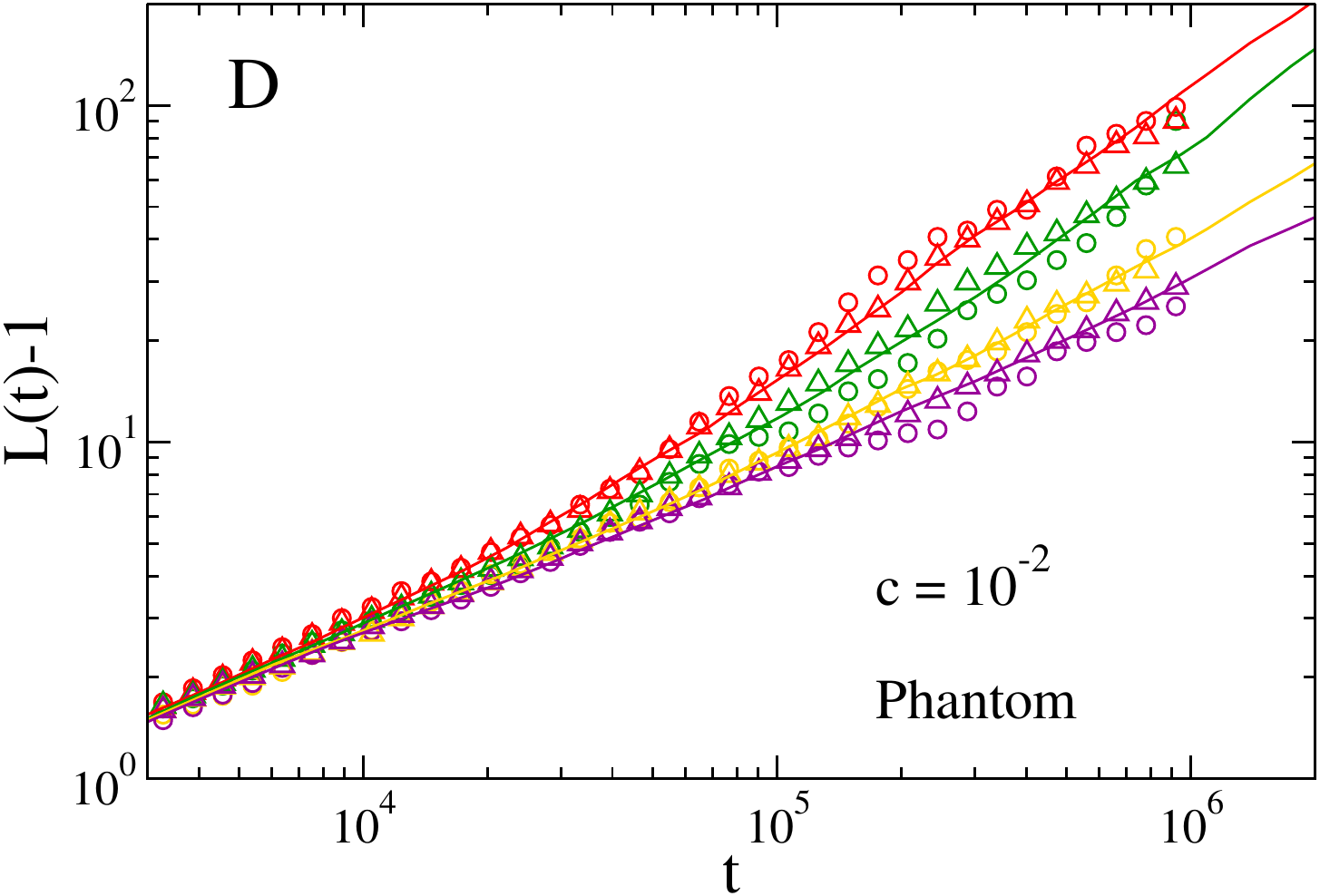}
\includegraphics[width=0.45\textwidth]{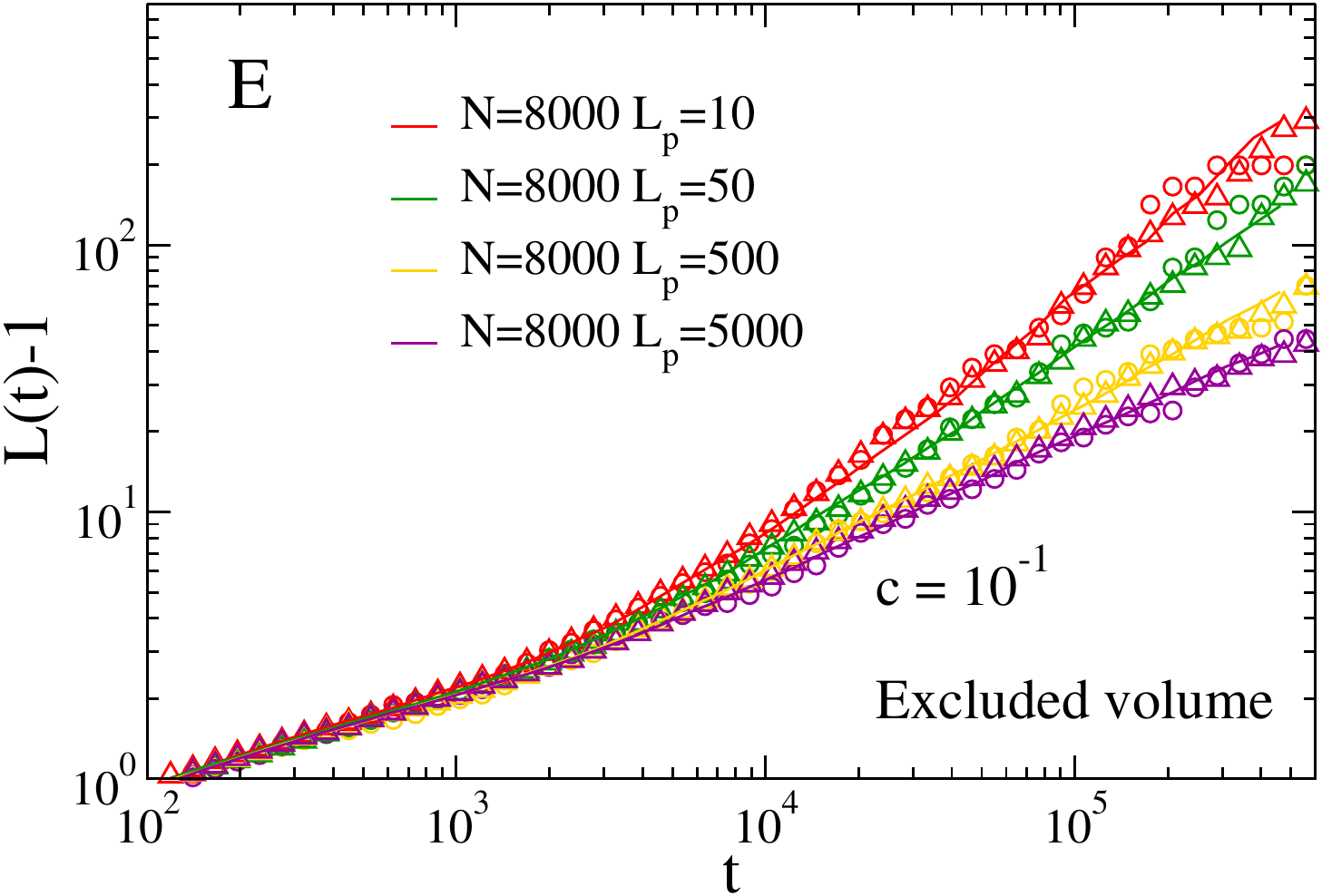}
\includegraphics[width=0.45\textwidth]{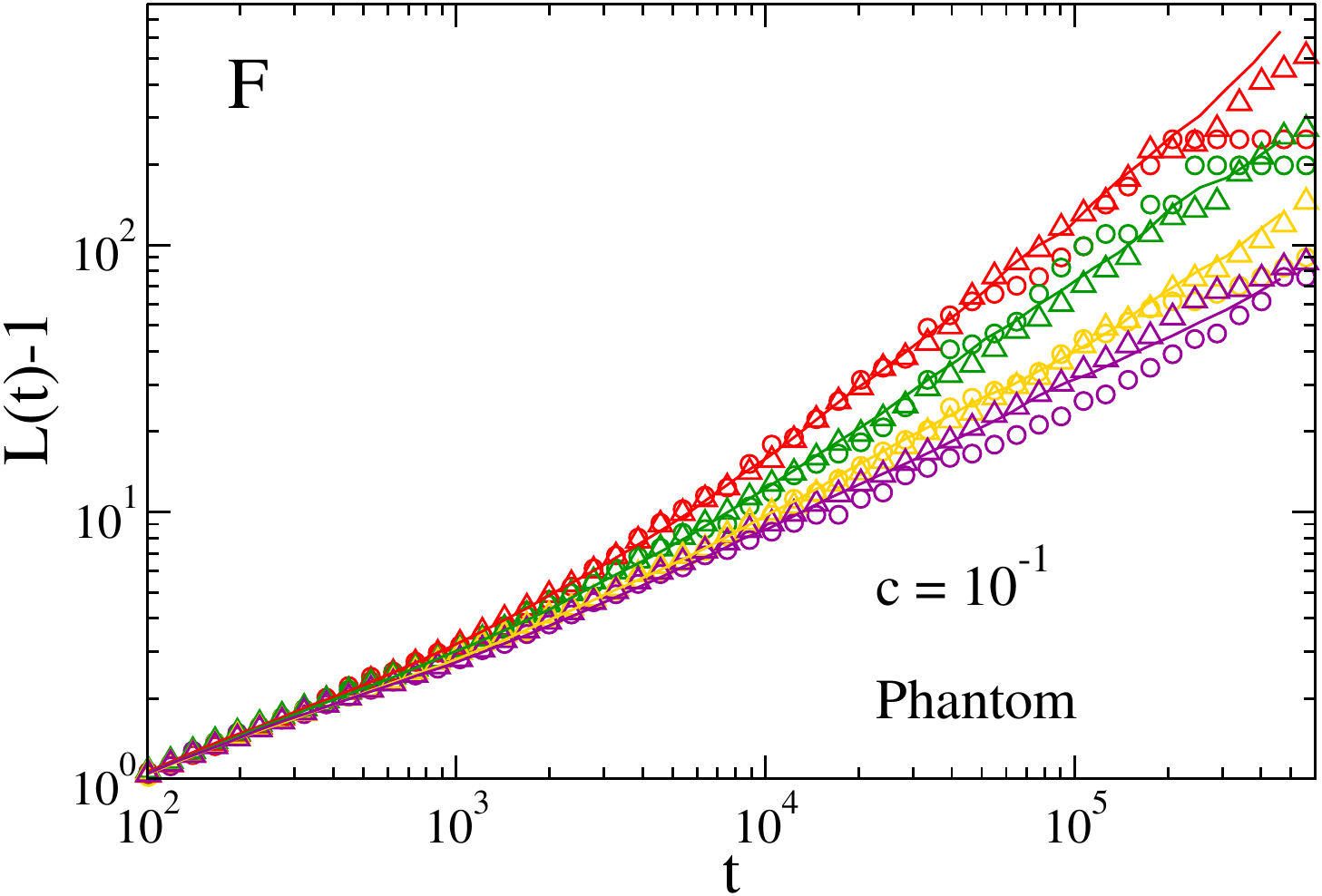}
\caption{Mean filament length for systems of different size (number of monomers) $N$, with excluded volume (A,C,E) and phantom (B,D,F) interactions, for different concentrations $c$. Circles: $N=1000$. Triangles: $N=4096$. Solid lines: $N=8000$. }
\label{fig:finite_size}
\end{figure}

In our simulation, the chosen system size is $N=8000$ monomers. Here, we address the possible presence of finite size effects in our simulations and show that the chosen system size is large enough to avoid these effects. In order to do so, we compare the results of the simulations reported in the paper for systems of $N=8000$ with those obtained for systems of $N=1000$ and $4096$ monomers. The results for the mean filament length $L(t)$ are reported in Fig.~\ref{fig:finite_size}, for persistence lengths $L_p=20$ (red lines/symbols), $L_p=100$ (green lines/symbols), $L_p=500$ (yellow lines/symbols), and $L_p=5000$ (purple lines/symbols). We have performed this analysis for all the concentrations considered in the main text $(c=10^{-3},10^{-2},10^{-1})$ and considering both systems with excluded volume interactions and phantom ones. One can see that the behavior of the simulated systems is, within the statistical noise, independent on the system size $N$. The only exception is observed for $N=1000$, where at times $t \gtrsim 10^5$ some systems (especially the ones with phantom filaments at the highest density $c=10^{-1}$, Fig.~\ref{fig:finite_size}F), reach a plateau in $L$. Analyzing the configurations revealed that this plateau is due to filaments looping with themselves across the periodic boundary conditions, an artifact that becomes much less likely for larger systems. However, as discussed in Sec.~\ref{sec:model}, the fraction of monomers which are part of loops is always $<6\%$ (and usually much smaller than this value) for system size $N=8000$, and thus we do not expect their presence to alter our results. From the analysis reported above, we thus conclude that finite size effects are negligible in the simulated systems of size $N=8000$.

%%%%%%%%%%%%%%%%%%%%%%%%%%%%%%%%%%%%%%%%%%%%%%%%%%%%%%%%%
\section{Analysis of filaments mean-squared displacement in the absence of assembly} \label{sec:msd}
%%%%%%%%%%%%%%%%%%%%%%%%%%%%%%%%%%%%%%%%%%%%%%%%%%%%%%%%%

Here we analyze the dynamics of individual filament ends depending on the presence or absence of excluded volume interactions. In our simulations, we observe that phantom polymers display faster growth than those with excluded volume interactions in the linear growth regime, \textit{i.e.}, they satisfy $L(t) \propto t$ but with a slightly larger prefactor (see Fig.~3). As discussed in the text, this could be attributed to a reduction of the prefactor of the mean-squared displacement (MSD) $x^2(t)$ of the filaments ends in the fluctuations-dominated regime, where $x^2(t) \propto t^{3/4}$. Here, we show that this is not the case by comparing the MSD of the reactive sites in a system of phantom polymers and in one with full excluded volume interactions, in the absence of annealing reaction.

\begin{figure}[h]
\centering
\includegraphics[width=0.55\textwidth]{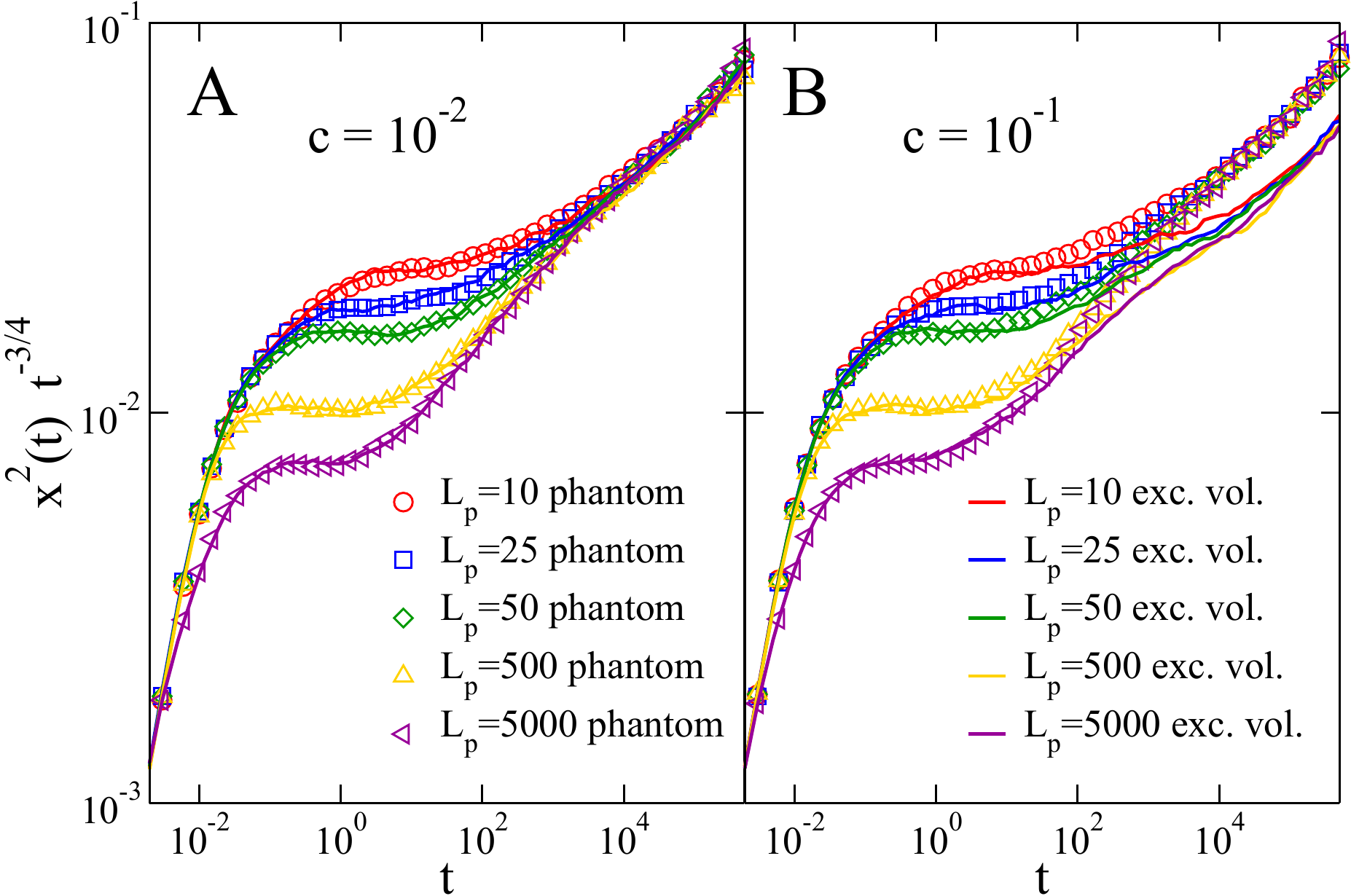}
\caption{Mean-squared displacement $x^2(t)$ of the end monomers of the filaments in the absence of annealing reactions. To highlight the bending-fluctuations regime, we have multiplied $x^2(t)$ by $t^{-3/4}$. (A): $c=10^{-2}$, $L=10.1$. Solid lines: excluded volume filaments. Symbols: phantom filaments. (B): $c=10^{-1}$, $L=10.2$. Lines and symbols are as in (A). The fluctuations-dominated regime [plateau $x^2(t) t^{-3/4} \simeq \text{const.}$] is unchanged when the excluded volume interactions are turned off.}
\label{fig:msd_ends}
\end{figure}

We simulate the excluded volume and phantom systems starting from the same initial configuration, with mean contour length $L\simeq 10$, for which a faster growth of the phantom polymers is already observed. For each simulation, we perform two independent realizations with different initial conditions and average the MSD data over these two realizations. In Fig.~\ref{fig:msd_ends} we show these data for $c=10^{-2}$ (A) and $c=10^{-1}$ (B). In order to highlight the bending-fluctuations regime, we plot $x^2(t) t^{-3/4}$ as a function of time instead of $x^2(t)$. In both panels, a plateau at intermediate times signals the presence of a $x(t) \propto t^{3/8}$ dynamical regime. Eventually, at longer times, the dynamics becomes diffusive, \textit{i.e.}, $x^2(t) \propto t$. We observe that for $c=10^{-2}$ (A), there is basically no difference between the phantom and excluded volume system. This leads us to speculate that the transient regime $L(t)\propto t^{0.4}$ observed in the presence of excluded volume interactions may be due to a slower filament relaxation following binding in the presence of excluded volume interactions. For $c=10^{-1}$ (B), again no difference is observed in the fluctuations-dominated regime, however the dynamics of the phantom filaments is faster at longer times as the filament can cross each other. Since a faster assembly of the phantom filaments is observed already for $L\simeq 10$ also for the lower density $c=10^{-2}$, as shown in Fig.~3, we conclude that this is not due to a faster dynamics of the reactive sites. 
 
%%%%%%%%%%%%%%%%%%%%%%%%%%%%%%%%%%%%%%%%%%%%%%%%%%%%%%%%%
\section{Filament length distribution}
%%%%%%%%%%%%%%%%%%%%%%%%%%%%%%%%%%%%%%%%%%%%%%%%%%%%%%%%%

In Sec.~\ref{sec:msd}, we have investigated whether the faster assembly observed for the phantom filaments was due to a faster dynamics of the reaction sites for phantom filaments, finding that this is not the case. Here, we investigate whether this faster dynamics could be due to differences in the filament length distribution $P(l)$ between the phantom filaments and those with excluded volume interactions, finding that this is not the case.  

\begin{figure}[ht]
\centering
\includegraphics[width=0.45\textwidth]{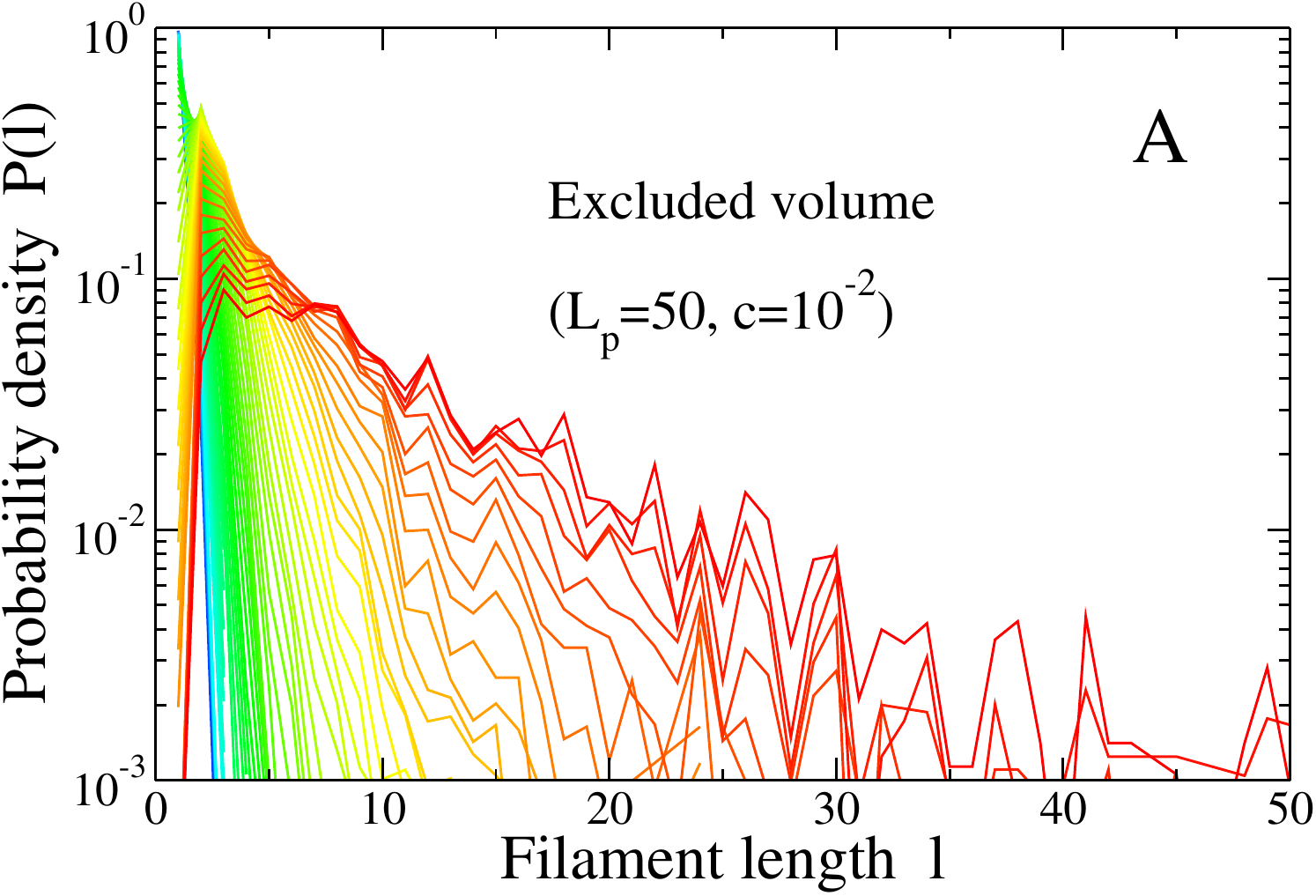}
\includegraphics[width=0.45\textwidth]{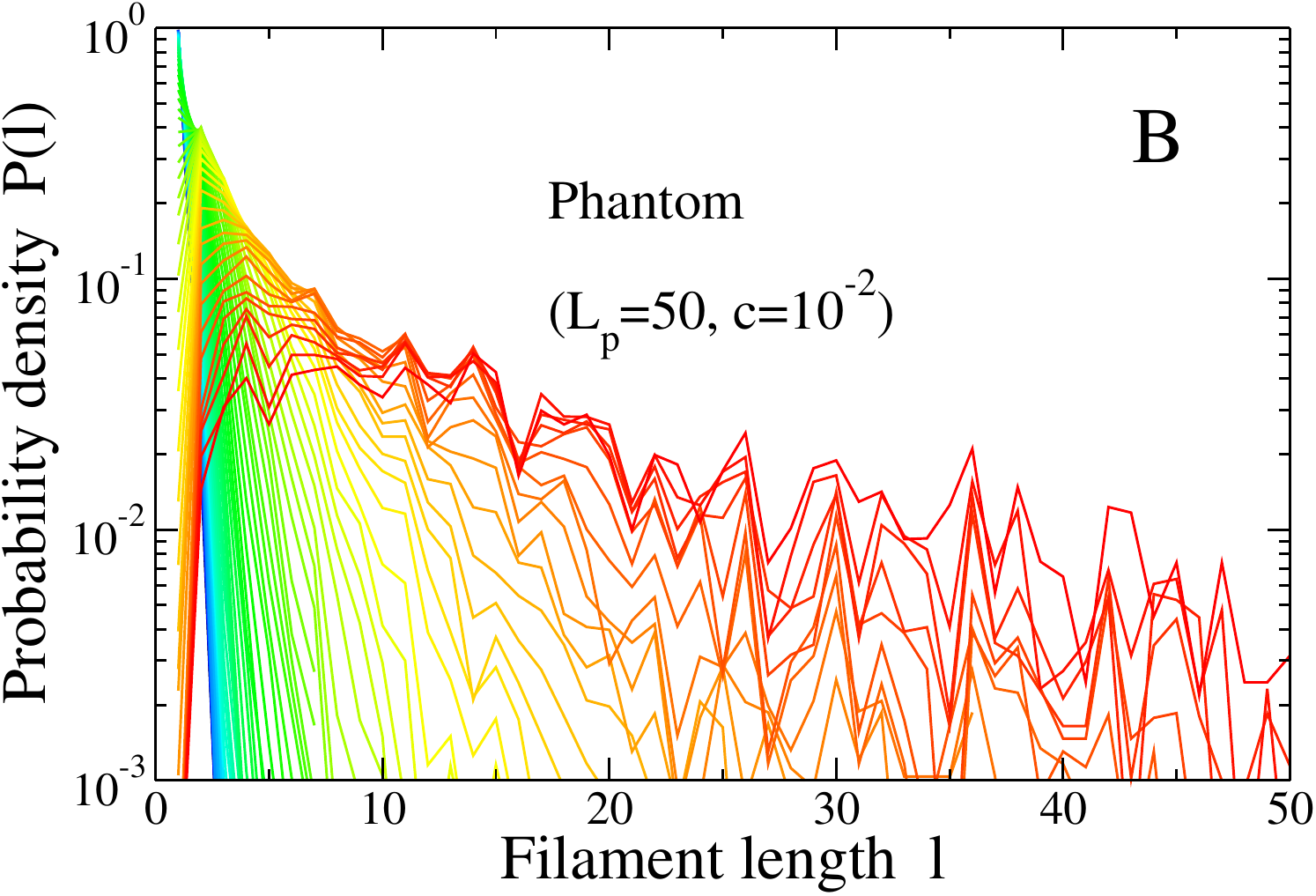}
\includegraphics[width=0.45\textwidth]{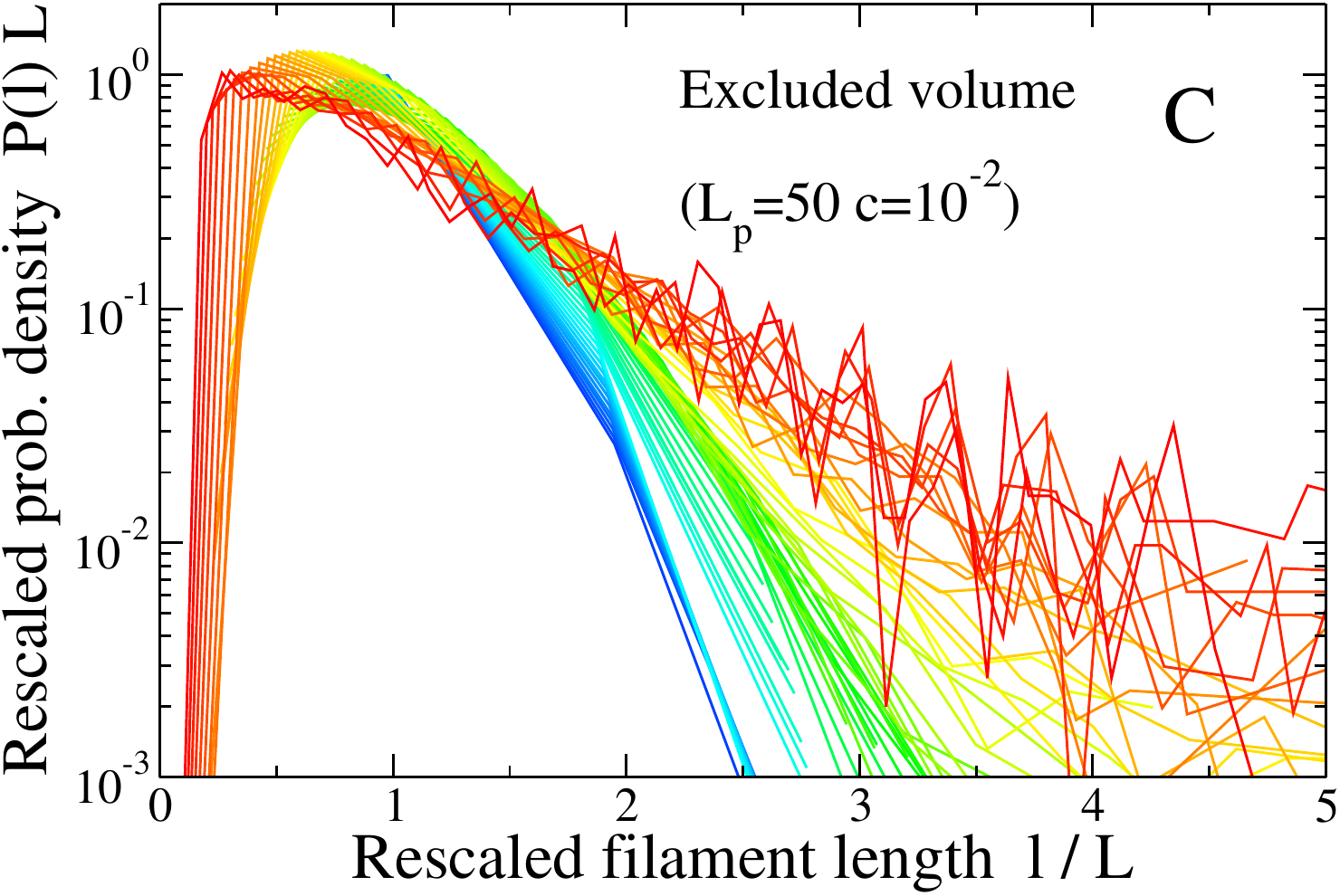}
\includegraphics[width=0.45\textwidth]{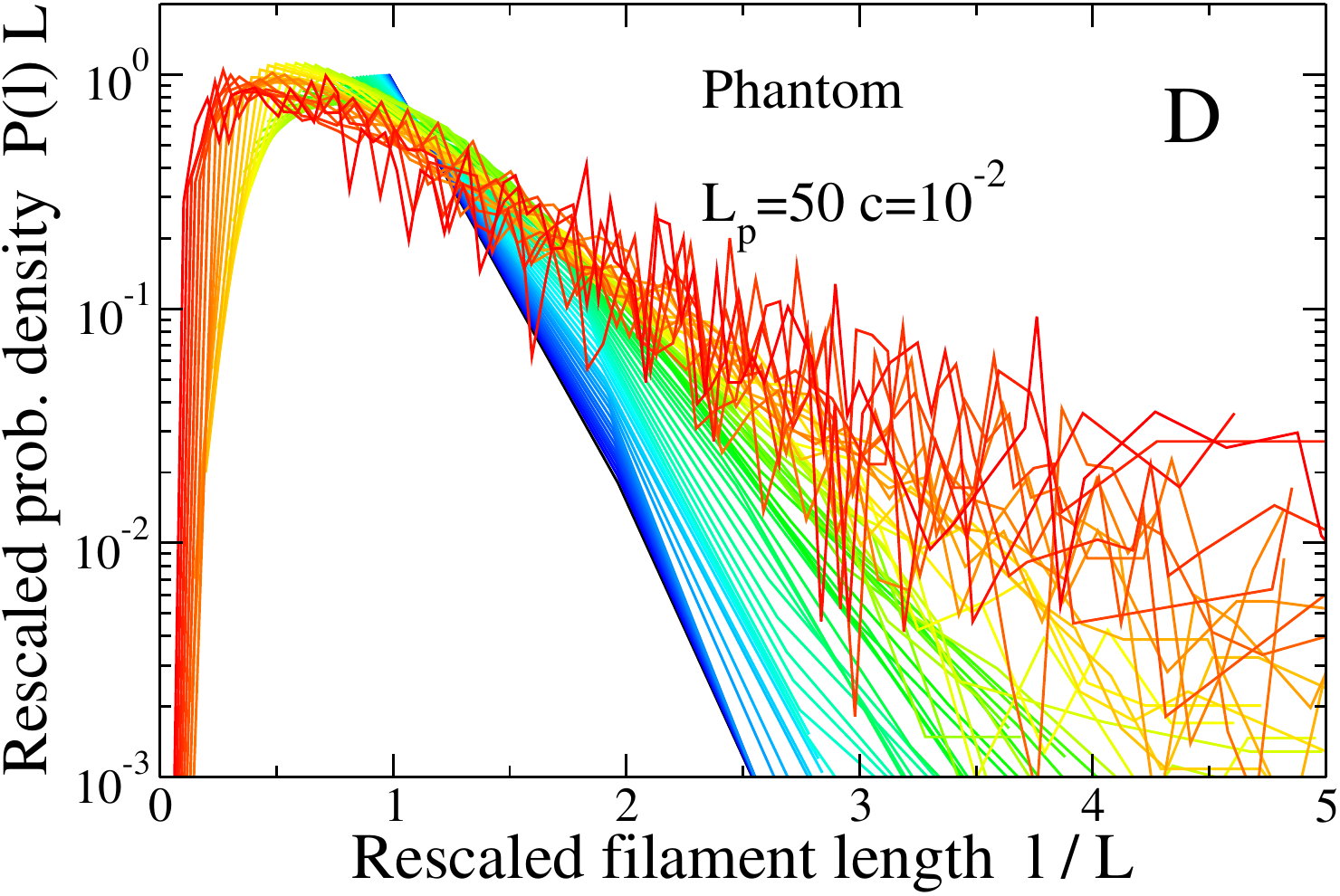}
\caption{Length distribution of the filaments $P(l)$ (probability density function) for $L_p=50$ and $c=10^{-2}$. (A): Filaments with excluded volume. (B): Phantom filaments. (C-D): Rescaled distributions, with $L$ the time-dependent mean length. Data are shown for times between $t_\text{initial}\simeq 0.8$ and $t_\text{final}\simeq 1.7 \times 10^5$, with the short-time curves shown in blue/green and the long-time ones shown in orange/red.}
\label{fig:pofl}
\end{figure}

We report in Fig.~\ref{fig:pofl}A--C the probability density function $P(l)$ for the systems $L_p=50, c=10^{-2}$, considering both filaments with excluded volume interaction (A,C) and phantom ones (B,D). Other systems show the same qualitative behavior. To reduce the statistical noise, we have defined 100 logarithmically spaced time windows and calculated the average $P(l)$ in each one of these windows. The data shown are for a single realization of the simulation and for times between $t_\text{initial}\simeq 0.8$ and $t_\text{final}\simeq 1.7 \times 10^5$, with the short-time curves shown in blue/green and the long-time ones shown in orange/red. Overall, one can see that the behavior of $P(l)$ is qualitatively very similar for the filaments with excluded volume and the phantom ones: At short times, it displays a simple exponential decay, whereas at longer times it develops a maximum close to the mean length $L$, followed by an exponential decay. Below, we describe in detail how $P(l)$ changes as the assembly proceeds, and compare excluded volume and phantom systems.

At short times (blue/green curves), when most of the reactions are comprised of monomer pairs reacting to form dimers, $P(l)$ decays exponentially. In this regime, the reaction rate is approximately constant and equal to the reaction rate of the monomers ($K \approx b^3 \tau^{-1}$, with $\tau \approx b^2 \zeta / k_B T$), which leads to the exponential decay~\cite{vandongen1984kinetics,sciortino2008growth}. As the assembly proceeds (orange/red curves), longer filaments form, whereas monomers (and then short filaments) are quickly depleted, as they are more mobile and thus react faster. This results in the emergence of a small-$l$ peak in the distribution, whereas for large $l$ the decay is still roughly exponential. 

In Fig.~\ref{fig:pofl}C--D we report the same data of Fig.~\ref{fig:pofl}A--B, rescaling $l$ by the mean length $L$. This is done to show how at long times, the exponential tails of the distributions for different times roughly fall on a master curve. Thus, the shape of the distribution remains roughly unchanged at long times. Finally, from this rescaling one can also see that the peak of the distribution at intermediate/long times is found at $l \simeq 0.4 L$. Thus, we conclude that there are generally very few filaments which are either much shorter or much longer than the average length $L$, justifying \textit{a posteriori} our scaling description in terms of a single typical length.

\begin{figure}[h]
\centering
\includegraphics[width=0.45\textwidth]{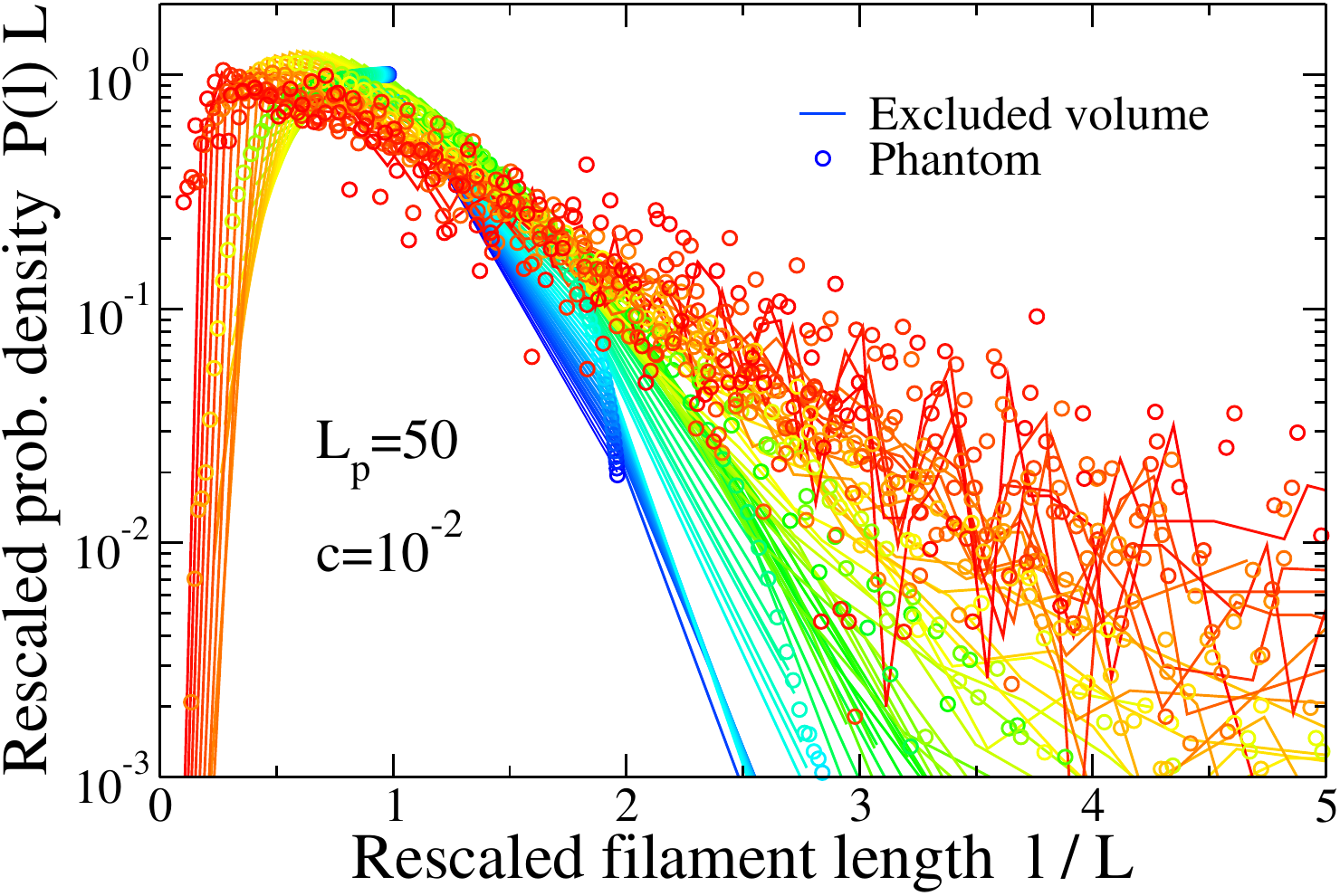}
\caption{Filament length distribution (probability density function) for $L_p=50$ and $c=10^{-2}$ as a function of the rescaled filament length $l/L$, with $L$ the time-dependent mean length. The data presented are the same as in Fig.~\ref{fig:pofl}A-B, here superimposed for comparison. Equal colors are assigned to equal time windows.}
\label{fig:pofl_cfr}
\end{figure}

By comparing Figs.~\ref{fig:pofl}A--B (or equivalently Figs.~\ref{fig:pofl}C--D), we observe that the qualitative shape of the distribution is very similar for excluded volume and phantom filaments. This is also highlighted in Fig.~\ref{fig:pofl_cfr}, where the length distributions for the two systems are superimposed onto each other (here equal colors are assigned to the same time windows). 

\begin{figure}[h]
\centering
\includegraphics[width=0.45\textwidth]{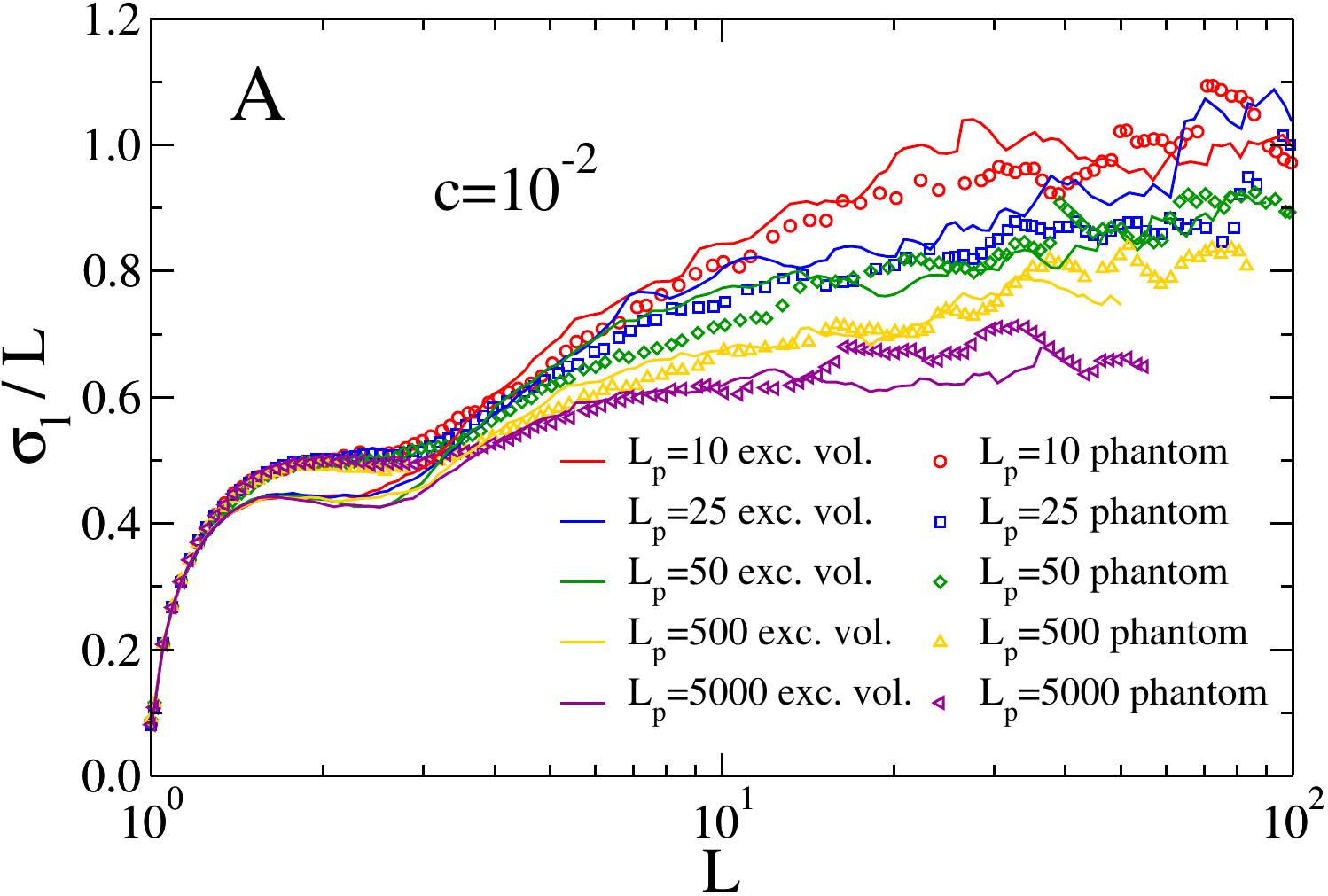}
\includegraphics[width=0.45\textwidth]{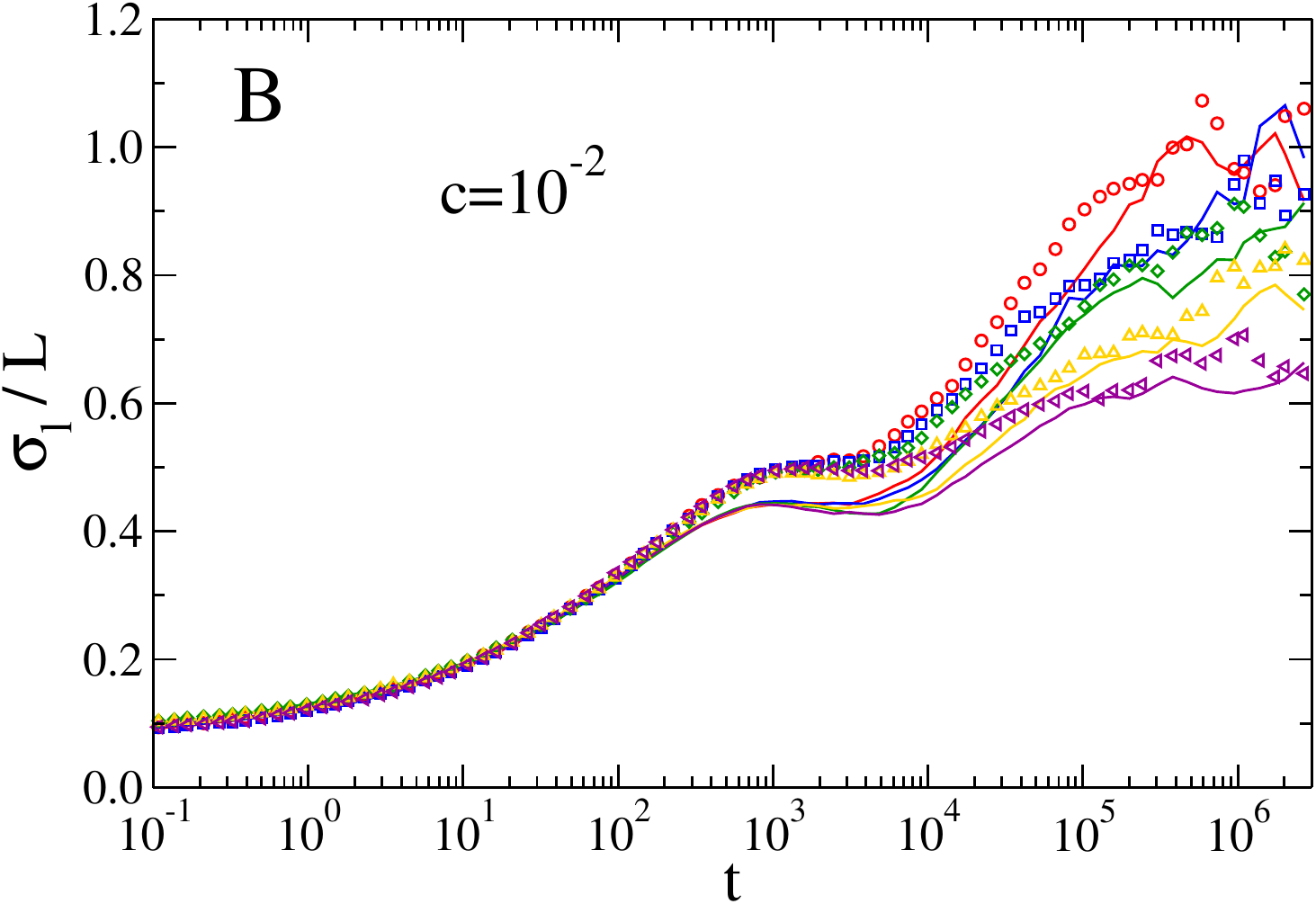}
\caption{Relative width of the filament length distribution, $\sigma_l/L$ (here $\sigma_l$ is the standard deviation of the distribution), as a function of the time-depenent mean filament length $L$ (A) and of time (B). Solid lines: filaments with excluded volume. Symbols: Phantom filaments.}
\label{fig:rel_std}
\end{figure}

To show that the shape of $P(l)$ is independent on the interaction potential for most values of $L$, we also show in Fig.~\ref{fig:rel_std}A the standard deviation $\sigma_l$ of the distribution, normalized by the mean length $L$, as a function of $L$. This quantity gives a measure of the relative width of the distribution. The choice of $L$ as the horizontal axis is here justified by the fact that phantom filaments react faster, and thus in order to compare $P(l)$ for excluded volume and phantom filaments it is preferable to take $L$ as a variable instead of time. For comparison, we also report in Fig.~\ref{fig:rel_std}B the same data as a function of time.  One can see in Fig.~\ref{fig:rel_std}A that, after an initial regime $(L \lesssim 4)$ in which the relative width depends on the interaction potential (phantom or excluded volume), the behavior of this quantity becomes interaction-independent. We can thus conclude that the slower reaction rate observed for the phantom system compared to the one with excluded volume is not due to differences in $P(l)$ for the two interaction potentials. Thus, as stated in the main text, we conclude that the faster assembly of phantom filaments to the inaccessibility of some reaction partners due to steric hindrance.

%%%%%%%%%%%%%%%%%%%%%%%%%%%%%%%%%%%%%%%%%%%%%%%%%%%%%%%%%
\section{Direct measurement of the reaction rate}
%%%%%%%%%%%%%%%%%%%%%%%%%%%%%%%%%%%%%%%%%%%%%%%%%%%%%%%%%

In the main text, we have shown that there is a crossover between a rigid-rod, sublinear growth regime [$L(t)\propto t^{1/2}$] and a fluctuations-driven, linear one [$L(t) \propto t$] by showing that the mean filament length data collapse on a master curve when $L$ is rescaled by the crossover length $L^*$ and time $t$ by the crossover time $t^*$. Here, present an alternative proof of the presence of this crossover by measuring the reaction rate $K$ directly from the reactions of same-length filaments.

\begin{figure}[h]
\centering
\includegraphics[width=0.45\textwidth]{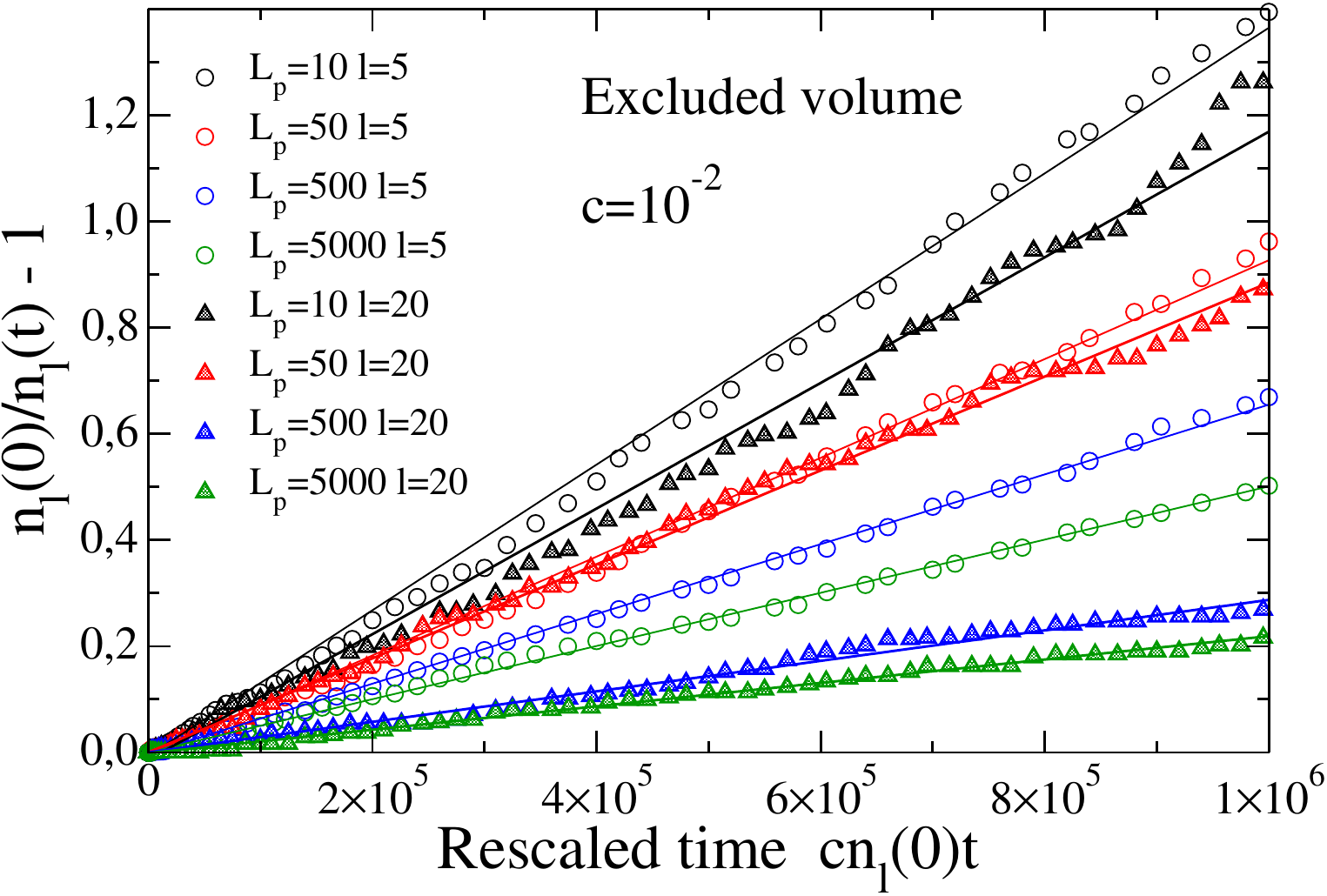}
\caption{Number of filaments of length $l$ at $t=0$, $n_l(0)$, divided by the same quantity at time $t$, $n_l(0)$. The data are presented as a  rescaled time $c n_l(0) t$. From Eq.~\eqref{eq:solution}], we expect this quantity to be a straight line with slope $K(l,l)$ at short rescaled times. Points are simulation data (shown here for a single realization), lines are linear fits to Eq.~\eqref{eq:solution}.} 
\label{fig:n_filaments}
\end{figure}

To do so, we have performed simulations in which the initial state is a well-mixed system of filaments with uniform length $l$. We have considered systems with concentrations $c=10^{-3}$ and $c=10^{-2}$, persistence lengths $L_p=10,50,500,5000$ and initial lengths $L(0)=l=5,10,20,50$, for both filaments with excluded volume and phantom filaments. For all systems, the total number of monomers is $N=10^4$.  Initially, the system is equilibrated until the rotational diffusion coefficient of the filaments, defined as $\langle \hat v^2(t)\rangle \equiv \langle | \hat v(t) - \hat v(0)|^2 \rangle$, with $\hat v \equiv {\mathbf R}_e/ |{\mathbf R}_e|$ the normalized end-to-end vector of the filament~\cite{doi1988theory}, reaches a plateau, ensuring that initial correlations in the orientation of the filaments have relaxed and the system has reached equilibrium. After this initial equilibration, we let the filaments assemble and track the number of filaments of length $l$ as a function of time, $n_l(t)$. Initially, we have $n_l(0)=N/l$ and the system only contains filaments of length $l$. Thus, we can assume that at short times only reactions of the type $l, l \rightarrow 2l$ take place, whence $n_l(t)$ is given by the following differential equation:

\begin{equation}
\dot n_l(t) = -K(l,l) c n_l^2(t),
\end{equation}

\noindent
which has the solution 

\begin{equation}
n_l(t) = \frac{n_l(0)}{1+K(l,l) c n_l(0) t},
\end{equation}

\noindent
which can be rewritten as a linear function:

\begin{equation}
\frac{n_l(0)}{n_l(t)} -1 = K(l,l) c n_l(0) t.
\label{eq:solution}
\end{equation}

\noindent
Thus, by plotting $n_l(0)/n_l(t)-1$ as a function of the rescaled time ${c n_L(0) t}$ and fitting it with a straight line, we can obtain $K(l,l)$. An example of this function and the fit is reported in Fig.~\ref{fig:n_filaments}, for $l=5,20$ and $c=10^{-2}$, for a system of filaments with excluded volume interactions. For each system, we perform three independent simulations starting from different initial conditions. The average reaction rate is then estimated as the average of the best fit values from the three realizations. We recall that our theory predicts that $K \approx b^3 \tau^{-1} (b/L)$ for $L \ll L^*$ (rigid rod regime), and $K \approx b^3 \tau^{-1} (L_p/b)^{-1/3}$ (fluctuations-dominated regime) for $L \gg L^*$, with $\tau \approx b^2 \zeta / k_B T$ the time a monomer takes to move by $b$ and $L^* \approx b (L_p/b)^{1/3}$ a crossover length. Thus, for $L \ll L^*$, we have $K \propto L^{-1}$, whereas $K$ is independent of $L$ for $L \gg L^*$. 

\begin{figure}[h]
\centering
\includegraphics[width=0.45\textwidth]{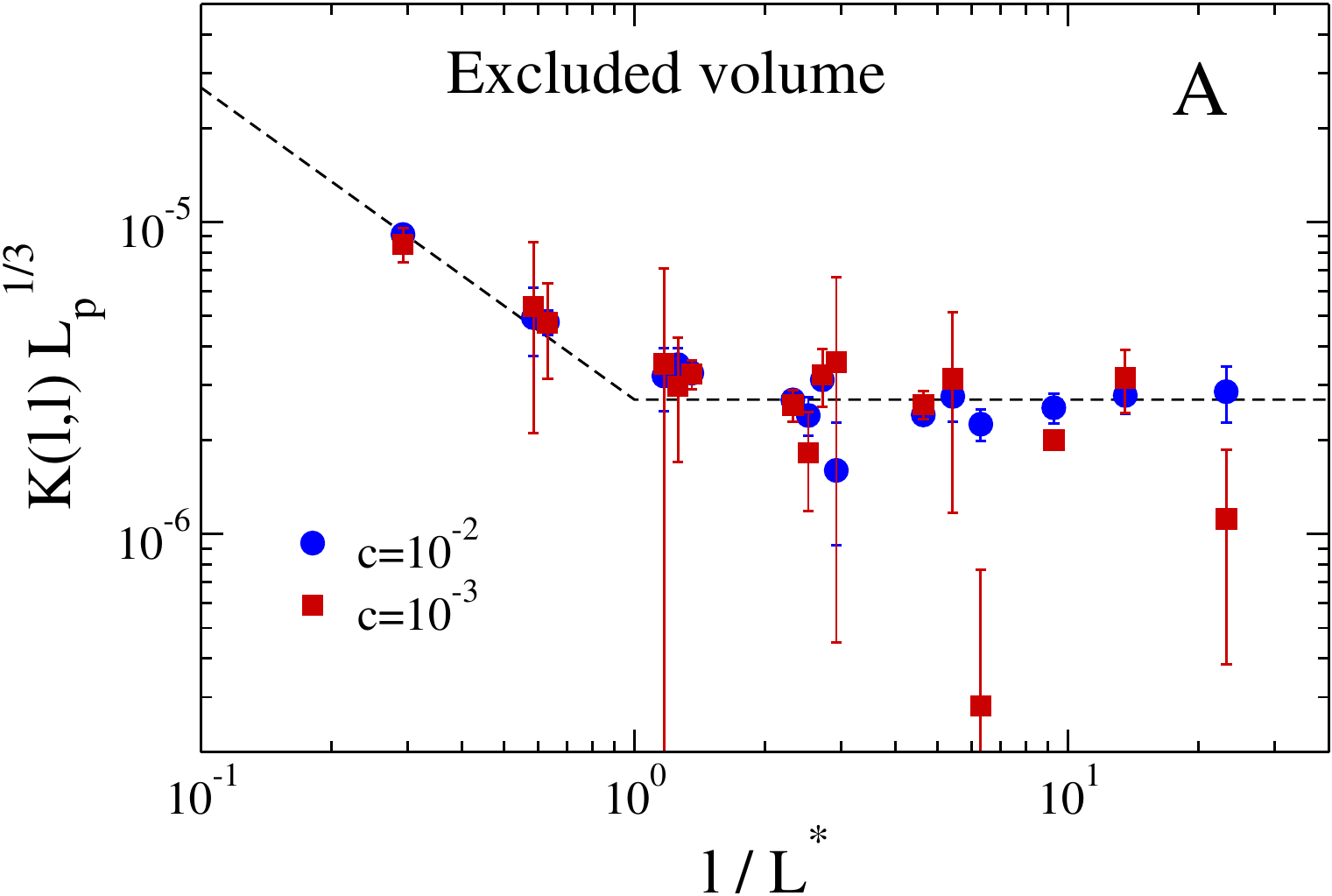}
\includegraphics[width=0.45\textwidth]{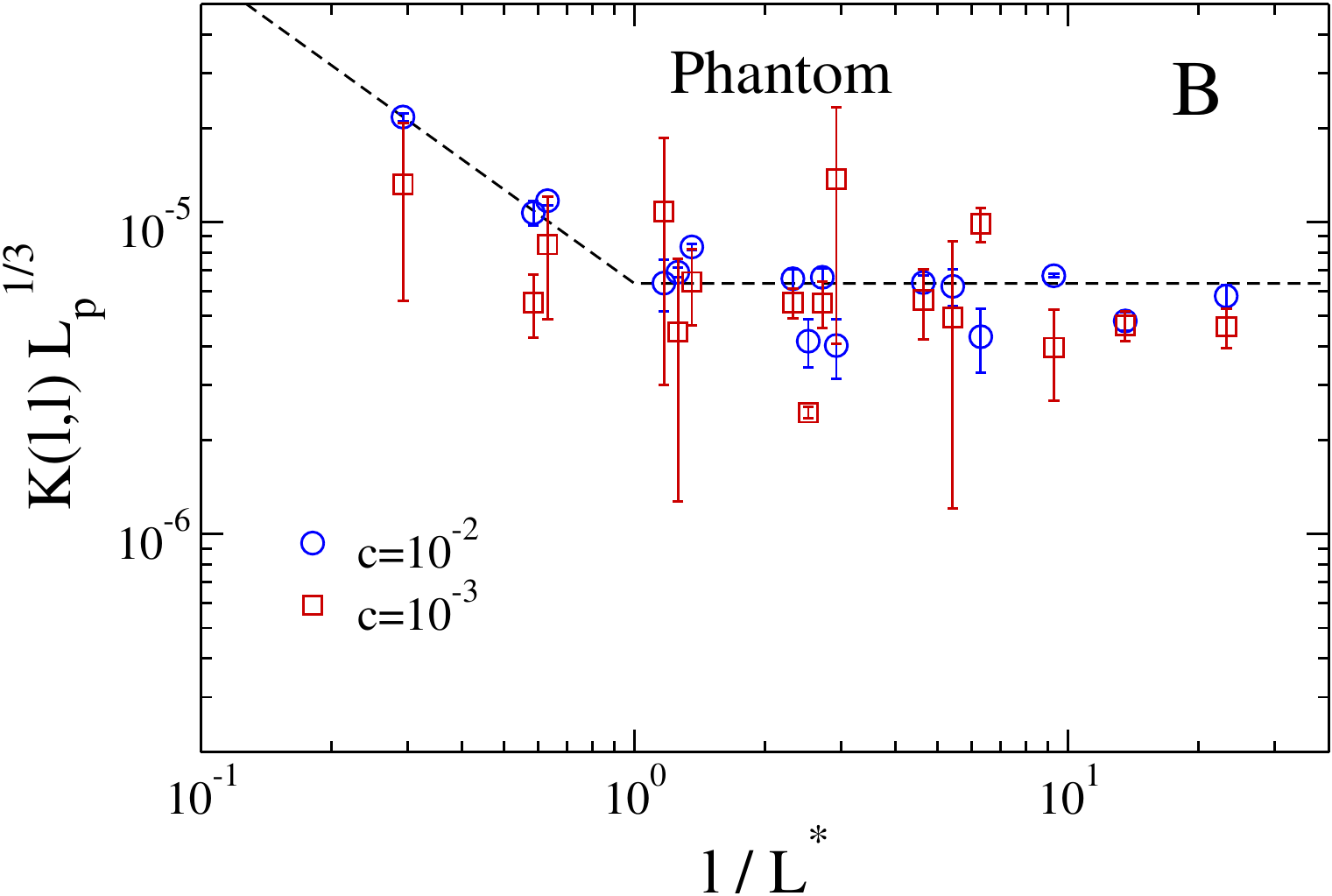}
\caption{Reaction rate $K(l,l)$ for a pair of filaments of same length $l$, rescaled by $L_p^{1/3}$, as a function of the rescaled length $l/L^*$, with $L^* = b (L_p/b)^{1/3}$. (A): Filaments with excluded volume. (B): Phantom filaments. Error bars represent standard deviations over three realizations. Dashed lines, reported to guide the eye, represent the predicted scaling behavior.}
\label{fig:k_rescaled}
\end{figure}

The values of $K(l,l)$ obtained from the fit are reported in Fig.~\ref{fig:k_rescaled}A--B for all the simulated systems. The dashed lines follow our scaling prediction and are drawn as a guide to the eye. One can see that, for both concentrations, the data points follow the scaling prediction within the statistical noise. In particular, we observe the expected crossover for $l/L^* \simeq 1$. The scaling behavior of $K(l,l)$ as a function of $l/L^*$ is the same for both filaments with excluded volume interactions (Fig.~\ref{fig:k_rescaled}A) and phantom filaments (Fig.~\ref{fig:k_rescaled}B), despite the rate being higher for the phantom filaments, as it was also observed and discussed in the main text. We note that some points for the lowest density ($c=10^{-3}$) apparently deviate more than the others from the scaling prediction; this is due to the low statistics associated to these systems, \textit{i.e.}, reactions are rarer at low $c$. 

In conclusion, the results of this analysis are in agreement with our scaling predictions. Additionally, they allow us to estimate $K$ directly, and to conclude that its value in the fluctuations-dominated regime is between $10^{-6}$ and $10^{-5}$ in simulation units.

%Rather, it is due to the fact that the filaments can cross each other, thus finding reaction partners more easily \cite{foffano2016dynamics,tran2022fragmentation}.

%%%%%%%%%%%%%%%%%%%%%%%%%%%%%%%%%%%%%%%%%%%%%%%%%%%%%%%%%
\section{Robustness of the assembly mechanism with respect to monomer addition}
%%%%%%%%%%%%%%%%%%%%%%%%%%%%%%%%%%%%%%%%%%%%%%%%%%%%%%%%%

In our simulations, the total number of monomers $N$ is kept fixed, \textit{i.e.}, there is not mechanism of monomer replenishment. As this mechanism can be present \textit{in vivo}, we partially address here its effects on the filament growth mechanism described in the main text. Starting from the more easily understood question of \emph{in vitro} experiments, we observe that our setup with a fixed total monomer concentration accurately captures the ones of
Refs.~\cite{falzone2013actin,kayser2012assembly,schepers2021multiscale,tran2022fragmentation}. By contrast, \textit{in vivo}
it is not entirely clear, depending on the system and process under consideration, that the monomer replenishment time is fast enough to significantly influence the assembly process. In addition, for systems like intermediate filaments, which mainly grow \textit{via} end-to-end annealing and are thus well described by our theoretical model, filament growth starting from monomeric units is still an open area of research~\cite{kirmse2007quantitative,portet2009vimentin, schween2022dual,tran2022fragmentation}. Moreover, in the presence of monomer replenishment, a filament clearance mechanism is needed in order to reach a steady state. Depending on the system under consideration, this mechanism could be severing of filaments, capping of their ends, or fragmentation. For example, it has been shown that vimentin filaments undergo fragmentation \textit{in vitro}, reaching an equilibrium length at long enough times~\cite{tran2022fragmentation}. The presence of a clearance mechanism would not influence the reaction rate $K$ directly, as this depends only on the length of the reacting filaments; however, it could influence it indirectly by changing the filament length distribution. Thus, if reasonable assumptions can be made about the clearance mechanism and how it influences the filament length distribution, it would still be possible to predict the assembly dynamics of the system by knowing the scaling of the reaction rate with filament length and solving the Smoluchowski equation with this rate.

\begin{figure}[t]
\centering
\includegraphics[width=0.45\textwidth]{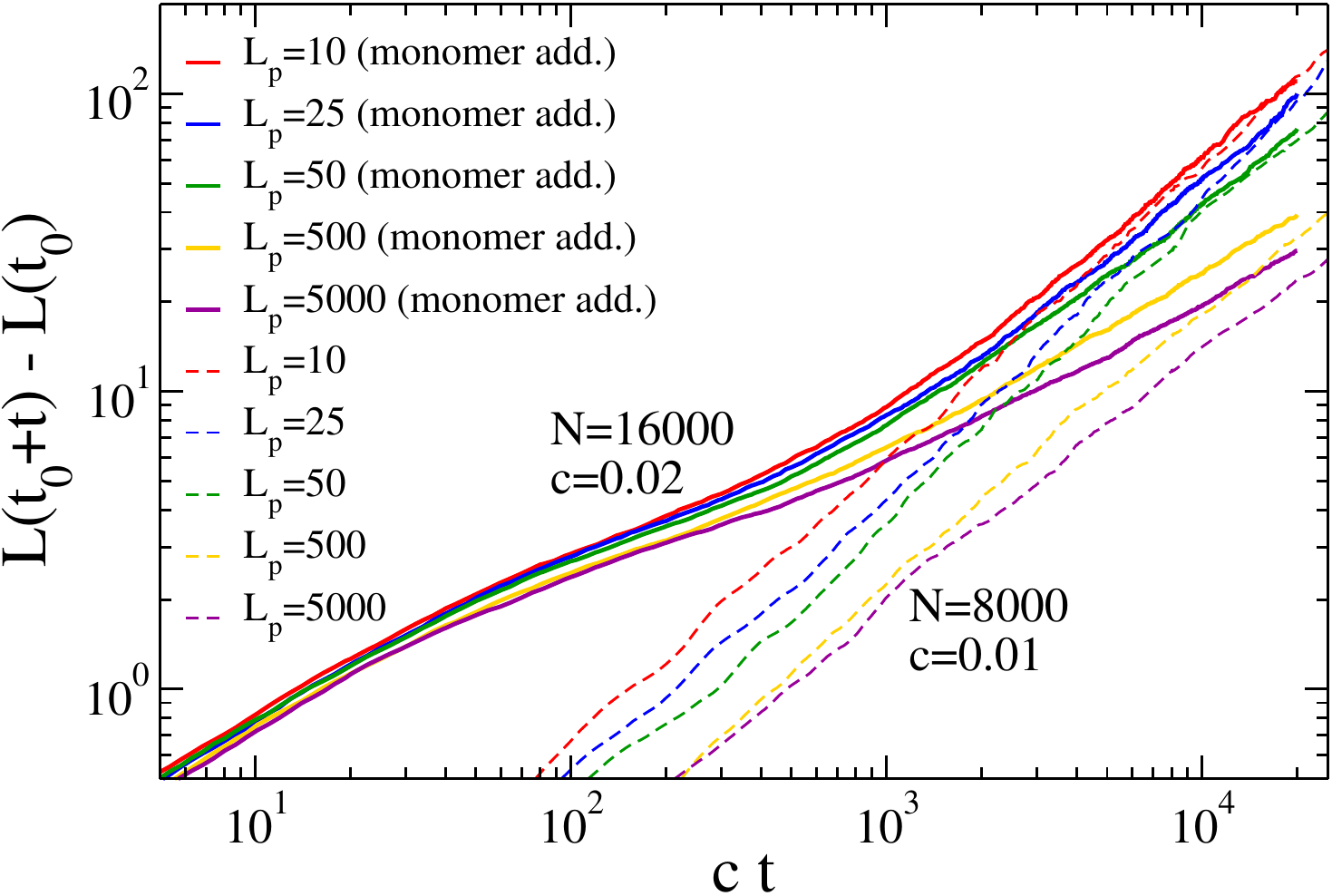}
\caption{Results of a ``replenishment experiment" in which monomers are added to systems with initial ($L_p$-dependent) mean length $L(t_0)$. Solid lines: $N$ monomers are added at $t_0$ to a system with initial length $L(t_0)$, after which the length increase $L(t+t_0)-L(t_0)$ is monitored. The total number of monomers is $N=16000$, with concentration $c=0.02$. Dashed lines: assembly without monomer addition ($N=8000$, $c=0.01$). To take into account the difference in concentration for the two systems, the data are reported as a function of the rescaled time $ct$.}
\label{fig:addmono}
\end{figure}

To test whether the predictions of our scaling theory are robust to the addition of monomers during the assembly process, we perform an \textit{in silico} ``replenishment experiment" that does not require us to make us any specific assumption on the filament clearance mechanism. To do so, we run the assembly of a systems with excluded volume interactions with $c=10^{-2}$ with different values of $L_p$ for a fixed amount of time $t_0$, so that the mean filament length reaches the ($L_p$-dependent) value $L(t_0)$. After this value is reached, $N=8000$ monomers are added to the system, so that the total number of monomers (and thus the concentration $c$) is doubled. After a short equilibration run to make sure that the monomers are uniformly distributed in the system, the assembly is then allowed to carry on for an additional time $t$. We then compare the quantity $L(t+t_0)-L(t_0)$ obtained after the addition of the monomers to the same quantity without monomer addition. We report the results of this comparison in Fig.~\ref{fig:addmono}. Since doubling $N$ means that $c$ is also doubled, and since we expect $L$ to be a function of $ct$ in all the regimes of interest (see main text), we plot the data as a function of the rescaled time $ct$. 

We observe that initially the assembly proceeds faster for the systems with the added monomers, due to the presence of fast-reacting monomers and short filaments. Then, as the assembly proceeds, these fast reacting components quickly form new filaments or are incorporated in the existing ones, following which the system resumes an assembly dynamics that is controlled by the concentration of these long filaments. This is highlighted in Fig.~\ref{fig:addmono} by the collapse of the curves for the system with added monomers (solid lines) on those for the systems without monomer addition (dashed lines), observed for $L_p < 500$. For the largest values of $L_p$ considered, $L_p=500$ and $5000$, a longer simulation would be needed to observe the collapse, but the same trend is observed. We thus conclude that, upon monomer addition, the monomers and small filaments are quickly incorporated in the pre-existing filaments, and the assembly kinetics remains controlled by the latter, whose length is of order of the typical length $L$.

%%%%%%%%%%%%%%%%%%%%%%%%%%%%%%%%%%%%%%%%%%%%%%%%%
\bibliography{bibliography.bib}